\documentclass[final,3p]{elsarticle}
\usepackage{amsmath,amscd}
\usepackage{bm,epsfig,ifthen}
\usepackage{amsfonts,amssymb}
\usepackage{colortbl}
\usepackage{pstricks}
\usepackage{mathtools}
\usepackage{float}
\usepackage{verbatimbox}
\usepackage{fancyvrb}

\DeclarePairedDelimiterX\braket[2]{\langle}{\rangle}{#1 \delimsize\vert #2}

\newcommand{\Frac}[2] {\frac{\textstyle #1} {\textstyle #2}}
\newcommand{\Int }    {\displaystyle \int}

\newcommand{\BigR}{{\rm I\kern-.27em R}}
\newcommand{\BigN}{{\rm I\kern-.25em N}}

\newcommand{\BigI}{{\rm I\kern-.25em I}}
\newcommand{\BigH}{{\rm I\kern-.25em H}}
\newcommand{\field}[1]{\mathbb{#1}}
\newcommand{\T}{\boldsymbol{\theta}}

\newcommand{\be}{\begin{equation}}
\newcommand{\de}{\end{equation}}

\begin{document}

\title{Computing non-equilibrium trajectories by a deep learning approach}
\author{Eric Simonnet}

\ead{eric.simonnet@inphyni.cnrs.fr}
\affiliation[1]{organization={INPHYNI, Universit\'e C\^ote d'Azur et CNRS, UMR 7010},
addressline={1361, route des Lucioles},
postcode={06560},
city={Valbonne},
country={France}}


\date{\today}
\begin{abstract}
Predicting the occurence of 
rare and extreme events in complex systems is a well-known problem in non-equilibrium
physics. These events can have huge impacts on human societies. New approaches have
emerged in the last ten years, which better estimate tail distributions. They 
often use large deviation concepts without the need to perform heavy direct ensemble simulations.
In particular, a well-known approach is to derive a minimum action principle and to 
find its minimizers.

The analysis of rare reactive events in non-equilibrium systems
without detailed balance is notoriously difficult either theoretically and computationally.
They are described in the limit of small noise 
by the Freidlin-Wentzell action. We propose here a new method
which minimizes the geometrical action instead using neural networks: it is
called {\it deep gMAM}. It relies
on a natural and simple machine-learning formulation of the classical gMAM approach.
We give a detailed description of the method as well as many examples. These include bimodal switches in complex stochastic (partial) differential equations, 
quasi-potential estimates, and extreme events in Burgers turbulence.
\end{abstract}
\maketitle

\section{Introduction}
Understanding rare and extreme events in complex models has become a cornerstone
of modern non-equilibrium physics. In many cases, the occurrence of such events has
strong impacts on our daily life and cannot be ignored despite their small probability of
occurrence. A well-known example is of course climate change and its impact at the regional 
scale. At the level of biological systems, dramatic phenomena can be tied to very small
unexpected changes, in biochemistry, molecular dynamics, 
with protein bindings/foldings for instance.
Remarkably, these events are predictable: they always conspire following the same
path associated with some well-defined probability. 
In fact, small fluctuations either random or deterministic can drive systems
to different unknown equilibria provided some energy barrier is reached. This is
the Arrhenius law. It is a very simple instance of a large deviation principle
(LDP) which expresses the property that a quantity (e.g. a probability) is behaving  as 
$-\epsilon \log \Pr \sim V$ when some parameter $\epsilon$ becomes small. 
Such LDP appears to be much more general than the original Arrhenius law and 
is not restrictive to the existence of a free energy potential $V$. In general situations
$V$ is then called the {\it quasi-potential} or rate function.
However, understanding non-equilibrium systems which do not have a detailed balance
turns out to be a very challenging task, not only theoretically, but computationally as well as.

The natural way to tackle these problems is large deviation theory as it provides a natural
and powerful framework for non-equilibrium statistical physics \cite{Touchette}.
In the context of metastability, the theory has been developed by Freidlin and Wentzell
in the 70s-80s \cite{FW} as a large deviation approach to perturbed dynamical systems
although these ideas were already known by the physicists with Onsager and Machlup \cite{Onsager}.
Noteworthy, there is another path to large deviations which is called
the Martin-Siggia-Rose-Jansen-de Dominicis (MSRJD) formalism developed in the
70s (\cite{Grafke} for a presentation).
Once a minimum action principle is obtained, it remains to identify the
minimizers. This is in general a very difficult optimisation problem as it involves
complex quantities related to the nontrivial nonlinear interactions with noise. 
It turns out that minimizing the Freidlin-Wentzell action directly is also very difficult
due to the long time it takes for the fluctuations to build up an optimal
path history. It translates mathematically as an optimisation problem 
in the large time limit which turns out to be ill-conditioned.  
A key step is therefore to identify an equivalent problem where time
has been reparametrized. It gives the so-called
{\it geometrical action} \cite{VdE_geo}.
This has led many authors
to consider the so-called gMAM approach bringing essential advantages over the original form.
How to minimize this action efficiently using neural networks and machine-learning techniques
is the subject of this work.

This work is organized as follow. We first recall fundamental notions of the Freidlin-Wentzell
large deviation theory and the concept of geometrical action in Section \ref{defs}.
We then describe in Section \ref{NN} 
how to adapt these problems to the machine-learning context and give
a detailed description of the deep gMAM approach and what it brings compared to the classical
one. Section \ref{exSDE} gives simple introductory examples of stochastic differential
equations (SDEs). It also illustrates as a byproduct a very simple way to
compute the quasi-potential for general SDEs.
Section \ref{exSPDE} considers more challenging 
examples of stochastic partial differential equations (SPDEs). 
We then illustrate the deep gMAM method 
in a very different context of extreme events in Burgers turbulence in Section \ref{burg1}.
We conclude in Section \ref{Concl}. Appendix \ref{FWzero} discusses a striking example of
a deterministic chaotic system and \ref{Julia} provides a simple snippet Julia code of
deep gMAM for the SDE case.


\section{Definitions and generalities}\label{defs}

\subsection{Freidlin-Wentzell large deviation theory}
We consider the following stochastic (partial) differential equation
\begin{equation}\label{des}
d u = F(u) ~dt + \sqrt{\epsilon} \sigma(u) dW_t,~u \in H
\end{equation}
where $\epsilon > 0, W_t$ is a Wiener process, $H$ is an ad-hoc Hilbert functional space, $u:(0,T) \times {\cal D} \to \field{R}^p: (t,{\bf x}) \mapsto
u(t,{\bf x})$ with ${\cal D} \subset \field{R}^d$ is the spatial domain and $u$ is a vector field
$(u_1,\cdots,u_p)$. For simplicity, we address the scalar fields only with $p=1$.
The correlation tensor (or diffusion matrix/operator) is
$$
\chi = \sigma \sigma^\star.
$$

The inner product is denoted $\langle \cdot,\cdot \rangle$ and the corresponding norm is $||u|| = (\langle u,u \rangle)^\frac12$.
We are interested in the probability $\Pr$ to have a time-$T$ trajectory (possibly with $T \to \infty$) connecting two
distinct states $a$ and $b$  when $\epsilon \to 0$. It is well-known from Friedlin-Wentzell theory \cite{FW} that it satisfies the large deviation principle (LDP)
\begin{equation}\label{LDP}
\lim_{\epsilon \to 0} -\epsilon \log \Pr   = \inf_{u \in {\cal C}}  S_T[u],
\end{equation}
where the set of trajectories starting from $a$ and ending to $b$ at time $T$ are denoted
\begin{equation}\label{BV}
{\cal C} \equiv {\cal C}_{a,b,T} = \{ u: (t,{\bf x}) \mapsto u(t,{\bf x})),~
u(0,{\bf x}) = a({\bf x}),u(T,{\bf x}) = b({\bf x}) \},
\end{equation}
and $S_T$ is the so-called Freidlin-Wentzell action:
\begin{equation}\label{FWA}
S_T[u] = \frac12 \Int_0^T \langle (\dot u - F(u),\chi^{-1} (\dot u - F(u)) \rangle~dt. 
\end{equation}
The argmin of the right-hand side (\ref{LDP}) is called in the literature an {\it instanton path}. In the
context where $a,b$ are local attractors for the deterministic dynamics ((\ref{des}) with $\epsilon=0$),
the (global) argmin solution is also often called a 
{\it non-equilibrium transition} or {\it maximum likelihood path}.
The minimum action viewed as a function  of $a$ and $b$ is 
called the {\it quasi-potential} \cite{FW,Cameron}, the definition in fact involves
another minimisation w.r.t. $T$:
\begin{equation}\label{qv}
	V: H \times H \to \field{R}, (a,b) \mapsto \inf_{T} \inf_{u \in {\cal C}_{a,b,T}} S_T[u].
\end{equation}

Strictly speaking, FW theorem applies for finite-dimensional systems only and is very generic
with mild conditions, e.g. continuity hypothesis. 
More specifically, 
the diffusion matrix $\chi$ must be uniformly non-degenerate:
$\langle p , \chi p \rangle \geq c ||p||^2, c > 0$ for a LDP to apply (see \cite{FW}, chap. 5.3). 
If not the BV constraints (\ref{BV}) cannot be satisfied.

The case of SPDEs is mathematically much more difficult to handle. 
A well-known historical example is 
a rigorous LDP proof for the Ginzburg-Landau 1-D PDE \cite{Jona}.
We thus consider all the following at a formal level only.
We will not write explicitly the dependency on ${\bf x}$ unless needed.
A more general formulation is to consider the minimum action problem as
\begin{equation}\label{Ham0}
\inf_{u \in {\cal C}} \Int_0^T L(u,\dot u)~dt ~{\rm with}~L(u,v) = \sup_{p} (\langle v,p \rangle - H(u,p)),
\end{equation}
where $H$ is the Hamiltonian. It can take different forms depending on the class of problems and
$L$ is its Legendre transform.
The advantage of this formulation is its generality, in particular when
the Lagrangian has no closed-form expression (see an example in \cite{FW}, chap. 5.3 and also \cite{Hey_VdE}). 
Moreover, one does not need to invert $\chi$.
In this work however, we focus on diffusion problems (\ref{des}) only, in which case the Hamiltonian
takes an explicit form 
\begin{equation}\label{FWH}
H(u,p) = \langle F(u),p \rangle + \frac12 \langle p,\chi p \rangle. 
\end{equation}
In the following, we propose an equivalent formulation of (\ref{Ham0}). Instead of 
considering the Lagrangian (\ref{FWA}): 
$L(u,\dot u) = \frac12 \langle \dot u - F(u),\chi^{-1} (\dot u - F(u)) \rangle$,
we consider the constrained problem:
\begin{equation}\label{Ham1}
\inf_{u \in {\cal C}} \frac12 \Int_0^T \langle p,\chi p \rangle~dt,~
	\chi p = \dot u - F(u).
\end{equation}
The general idea of the proposed deep gMAM method, is to solve (\ref{Ham1}) by a penalisation method.
We can therefore handle more complicated situations where $\chi^{-1}$ cannot be expressed easily by enriching the cost functional by an additional penalty term. We recall that the notation $\chi p$ means the operator $\chi$ applied to the field $p$. This formulation is very natural, a similar
approach can also be found in \cite{Tobiasns3d}.
\\\\
We will provide a highly nontrivial PDE example with a complicated $\chi$ in section \ref{burg1}.
We next describe the so-called geometrical action.

\subsection{Geometrical action}\label{gmam}
It is often the case that one is interested in minimizing $S_T$ w.r.t. $T$ as well. It is in general equivalent to consider the limit $T \to \infty$.
This situation typically happens in the context of transitions from one equilibrium state 
(or loosely speaking a local attractor) to another,
the transition takes an infinite amount of time to escape from the starting equilibrium.
The minimisation problem is therefore
$$
\inf_{T > 0} \inf_{{u \in {\cal C}}} \frac12 \Int_0^T ||\dot u - F(u)||_{\chi^{-1}}^2~dt,
$$
with $||f||_{\chi^{-1}} \equiv \langle f,\chi^{-1} f \rangle$ the induced covariance norm, see (\ref{FWA}).
In order to avoid the double minimisation problem, it is possible to derive an equivalent formulation
which removes the time constraint as shown by \cite{Hey_VdE} (see also \cite{FW98}).
The idea is the following. Using $||u||^2 + ||v||^2 \geq 2 ||u||~ ||v||$ with equality when $||u||=||v||$, 
one has $\Frac12 \Int_0^T ||\dot u - F(u)||_{\chi^{-1}}^2~dt
\geq \Int_0^T \left(||\dot u||_{\chi^{-1}} ||F(u)||_{\chi^{-1}} - \langle \dot u, F(u) \rangle_{\chi^{-1}} \right)~dt$.
The remarkable observation is that this inequality is indeed an equality. One can choose a reparametrization
of time say $\gamma(t) \geq 0$ such that $\gamma(0) = 0$ and $\gamma(T) = T$, the change of variable
$t = \gamma(\tau)$ yields $\Int_0^T \left( 
||\dot u||_{\chi^{-1}} ||F(u)||_{\chi^{-1}} - \langle \dot u, F(u) \rangle_{\chi^{-1}} \right) dt =
\Int_0^T \left( ||u'||_{\chi^{-1}} ||F(u)||_{\chi^{-1}} - \langle u', F(u) \rangle_{\chi^{-1}} \right)~d\tau$,
with $u' = du/d\tau$. The trick is now to choose $\tau$ such that $||u'||_{\chi^{-1}}=||F(u)||_{\chi^{-1}}$.
This is always possible, by taking $\gamma' = ||\dot u||_{\chi^{-1}}/||F(u)||_{\chi^{-1}}$.
The minimisation problem is equivalent to (changing the notation say $\dot u  = du/d\tau$)
\begin{equation}\label{geoa}
\inf_{u \in {\cal G}} \Int_0^1 \left( ||\dot u||_{\chi^{-1}} 
||F(u)||_{\chi^{-1}} - \langle \dot u, F(u) \rangle_{\chi^{-1}} \right)~d\tau,~
{\cal G} \equiv \{ u:\tau \to u(\tau), u(0) = a,u(1) = b \}.
\end{equation}
It is therefore the location on the curve which prevails, the way it is parametrized does not matter.
A more rigorous proof can be found in \cite{Hey_VdE}.
\\\\
In practice, when $\chi$ has a complicated form, ie when $\chi^{-1}$ has no explicit formulation, 
we just formulate the same problem like in (\ref{Ham1})
avoiding $\chi^{-1}$, namely solving
the constraint problem
\begin{equation}\label{finalgeoa}
\inf_{u \in {\cal G}} \Int_0^1 \left( ||v||_{\chi} 
||w||_{\chi} - \langle v,w \rangle_{\chi} \right)~d\tau,~
\chi v = \dot u,{\rm and}~\chi w = F(u).
\end{equation}
and its penalisation (quadratic) formulation:
\begin{equation}\label{pengeoa}
\inf_{u \in {\cal G},v,w} \Int_0^1 \left( ||v||_{\chi} 
||w||_{\chi} - \langle v,w \rangle_{\chi} \right)~d\tau
+ \gamma_v \Int_0^1 ||\chi v - \dot u||^2~d\tau + \gamma_w \Int_0^1 ||\chi w - F(u)||^2~d\tau.
\end{equation}
The penalisation parameters $\gamma_v,\gamma_w \gg 1$ must be large enough but at the same time
not too large to avoid dealing with an ill-conditioned functional.
The norm involved in the geometrical action is the $\chi$-norm, by contrast with the two
constraints on $(v,w)$ involving some user-defined one, e.g. an Euclidian weighted norm.
The treatment of (\ref{finalgeoa}) using (\ref{pengeoa}) is not the only possibility, other strategies
are possible, e.g. augmented Lagrangian methods (see \cite{Aug} in the machine-learning context 
applied to elliptic and eigenvalue problems, and \cite{Tobiasns3d} 
using some classical approach). One can also consider
the minimax original problem (\ref{Ham0}) using adversarial networks.
In the context of machine-learning however, (\ref{pengeoa}) 
appears to be the simplest strategy giving very good results. This is explained in details
in the next section \ref{NN}.


\section{Deep gMAM method}\label{NN}
We describe here in more details how to use Neural Networks (NNs) for solving (\ref{pengeoa}).
The general idea is to parametrize the argmin solution of (\ref{pengeoa}) by some NNs. 
In doing so, we already make a choice of working into the physical space-time domain $(t,{\bf x}) \in
(0,1) \times {\cal D}$. Due to that, convolutional neural networks 
are not appropriated in this context. It amounts
to the fact that one has access to a macroscopic description of the action and takes full advantage of it.
The broad philosophy of the proposed approach is therefore the same than for the so-called 
physics-informed neural networks (PINNs), \cite{pinns,dgm,NeuralPDE}
except that the system to solve is more involved with possibly much more complicated constraints. 
We will discuss the technical aspects below in the subsections
\ref{NN0}--\ref{NN5} and summarize what it can bring in the last subsection \ref{NN6}.

\subsection{Neural Network parametrization and cost functional}\label{NN0}
We consider here one of the simplest neural architecture, the so-called fully-connected feedforward NNs.
Let ${\cal N}: \field{R}^{d+1} \times \field{R}^{n_\theta} \to \field{R}, (t,{\bf x},\T)
\mapsto {\cal N}(t,{\bf x};\T)$, where $\T$ is the vector of $n_\theta$ parameters. The function ${\cal N}$
is defined as
$$
{\cal N}(t,{\bf x};\T) = {\cal N}_{\rm out} \circ {\cal N}_L \cdots {\cal N}_1 \circ {\cal N}_{\rm in}(t,{\bf x}),
$$
where $\circ$ is the composition and ${\cal N}_k({\bf y}) = \sigma_k(W_k {\bf y} + {\bf b}_k)$.
The parameters ${\bf b}_k \in \field{R}^{l_k}$ are called NN biases 
and the $(l_k \times c_k)$ matrices $W_k$ are called NN weights. The activation
functions $\sigma_k$ must be nonlinear. In the examples discussed later, the dimensions are chosen
independent of $k$ with $l_k=c_k = c$. The number of hidden layers $L$ is the depth, whereas $c$ is the network capacity. In the following, we consider only
{\it swish} activation functions \cite{swish}.
They correspond to some non-monotonic version of the classical RELU functions $\max(0,x)$,
with $\sigma(x) = x/(1+{\rm e}^{-x})$. Other choices are of course
possible, such as RELU or $\tanh$ but in practice the swish nonlinearity gives very good results and at the
same time smoother representations. The last layer ${\cal N}_{\rm out}$ is in general linear. 
The depth $L$ and $c$
are important parameters. The expressivity of ${\cal N}$ becomes better as $L$ and $c$ increases but at the
same time, it becomes harder to train, ie finding ad-hoc optimal values for $\T$.
\\\\
As stated above, we now replace the main and auxiliary fields  $(u,v,w)$ in (\ref{pengeoa}) by their NN parametrizations $u \to {\cal N}_u$, 
$v \to {\cal N}_v$ and $w \to {\cal N}_w$ with parameters 
$\T_u,\T_v,\T_w$. The cost functional is then minimized with respect to these parameters: 
we have thus performed a nonlinear projection onto the neural network space.
This is indeed a very general methodology in machine-learning. The cost functional is written as
\begin{equation}\label{generalC}
{\cal C}[\T_u,\T_v,\T_w] = \gamma_g {\cal C}_g[\T_v,\T_w] + \gamma_{\rm arc} 
{\cal C}_{\rm arc}[\T_u] + {\cal C}_{\rm constraints} + (\gamma_{\rm bcs} {\cal C}_{\rm BCs.}).
\end{equation}
The first term corresponds to the geometrical action in its continuous form:
\begin{equation}\label{NNAg}
{\cal C}_g[\T_v,\T_w] = \Int_0^1 \left(||{\cal N}_v||_{\chi} ||{\cal N}_v||_{\chi} - \langle {\cal N}_v,{\cal N}_w
\rangle_\chi \right)~d\tau, ~||{\cal N}||_\chi \equiv \left(\Int_{\cal D} {\cal N}(t,{\bf x})~
(\chi {\cal N})(t,{\bf x})~d{\bf x} \right)^\frac12.
\end{equation}
An important issue is to insure that when $||\cdot ||_\chi$ is discretized, one still has
the properties of a norm, e.g. definite positive and Cauchy-Schwarz inequality. 
It is in general trivial (e.g. $L^2 \to l^2$) 
but it can be more tricky if $\chi$ is complicated:  
a nontrivial case is discussed in subsection \ref{NN5}.
We now describe the other functionals in the next subsections.

\subsection{Boundary value ansatz}\label{NN1}
The cost functional (\ref{generalC}) must take into account the boundary value (BV) problem, namely
that ${\cal N}_u(0,{\bf x}) = a({\bf x})$ and ${\cal N}_u(1,{\bf x}) = b({\bf x})$ where $a,b$ are given.
There are two strategies: either imposing these constraints explicitly by penalisation or considering some 
ansatz for ${\cal N}_u$ which automatically include them. The second choice means that
one uses the general ansatz
\begin{equation}\label{ansatz}
{\cal U}(t,{\bf x};\T_u) = \Lambda_a(t) a({\bf x}) + \Lambda_b(t) b({\bf x}) + 
\Lambda_u(t) {\cal N}_u(t,{\bf x};\T_u),
\end{equation}
where $\Lambda_a(0) = 1$, $\Lambda_a(1) = 0$, $\Lambda_b(0) = 0$, $\Lambda_b(1) = 1$, $\Lambda_u(0)=
\Lambda_u(1) = 0$. In addition, the zeros of the functions $\Lambda$ must be only those required
at the boundaries $\tau = 0, \tau = 1$.

We then replace ${\cal N}_u$ in (\ref{generalC}) by ${\cal U}$ instead
whenever it is explicitly needed.
In practice, we use $\Lambda_a(t) = 1-t$, $\Lambda_b(t) = t$ and $\Lambda_u(t) = t(1-t)$ mimicking a
double Taylor expansion. We do not claim for optimal decision here, as many other choices would work. This
one gives very good results in all the situations we have met. 
It is preferred over the first approach (BV penalization) when it is difficult
for the system to relax on $a$ and $b$. 

\subsection{Arclength condition}\label{NN2}
Although the geometrical action does not depend on the time parametrization chosen due to its homogeneity, it is important in practice to restrict the problem. As a matter of fact, in the cost functional landscape, there is an infinity
of possible solutions, each having its own parametrization. Fixing an arclength condition, say
$||\dot u|| = c, \forall s \in [0,1]$ is just a matter of convenience. But more importantly, it stabilizes
the gradient search preventing the NNs to drift towards ill-conditioned parametrizations, especially in the
context of PDEs. The use of a small penalisation parameter $\gamma_{\rm arc} \ll 1$ 
is enough to prevent NNs to explore extreme 
landscape regions. The penalty constraint after straightforward algebra is
\begin{equation}\label{arc}
{\cal C}_{\rm arc}[\T_u] = \Int_0^1 ||\dot {\cal U}||^2~ds - \left( \Int_0^1 ||\dot {\cal U}||~ds \right)^2 \geq 0,
\end{equation}
where ${\cal U}$ takes the form (\ref{ansatz}). Note that the norm used is user-defined rather than
the actual $\chi$ or $\chi^{-1}$ norm.  
\subsection{Additional constraints}\label{NN4}
We now discuss the constraints: the ones associated with the auxiliary fields $v,w$ and external ones
which might be needed in some specific cases when looking for minimizers in subspaces of ${\cal C}$.
They are, from (\ref{pengeoa}),
\begin{equation}\label{aux}
{\cal C}_{v,w} = \gamma_v \Int_0^1 ||\chi {\cal N}_v - \dot {\cal U}||^2~d\tau + 
\gamma_w \Int_0^1 ||\chi {\cal N}_w - F({\cal U})||^2~d\tau.
\end{equation}
The spatial derivatives involved in $F$ (or possibly in the operator $\chi$) 
and the derivative $\dot {\cal U}$ can be obtained
either using an efficient automatic differentiation (AD) algorithm or using finite-difference schemes with
a very small increment (consistent with the scheme order and the machine precision, see PINNs methods).
\\\\
Other constraints might be needed. 
For instance, in some situations (see examples \ref{ngSDEs} and \ref{ng1PDEs},\ref{ng2PDEs}), one must fix
a total length $L$ for the instanton path, in this case, one uses
\begin{equation}\label{instleng}
{\cal C}_L[\T_u] = \gamma_L \left( \Int_0^1 || \dot {\cal U} || ds - L \right)^2,
\end{equation}
where $L$ is fixed by the user. Other typical situations are problems having (discrete) symmetry groups.

One can easily constrain the NNs to satisfy or break some symmetry. We do not write the 
penalty terms as they are problem-dependent (see an example (\ref{symmbreak})).
This is a very useful technique to detect candidates for global minimizers 
(see subsection \ref{ng2PDEs}).

\subsection{Boundary conditions (PDEs)}\label{NN3}
For PDE problems, one must impose some boundary conditions on $]0,1[ \times \partial {\cal D}$.
Like in subsection \ref{NN1} there are two choices: either penalizing the boundary constraints 
or finding some ad-hoc 
ansatz. Both strategies often yield equivalent results unless more precision is required. We discuss here
only the case of periodic boundary conditions and propose a new approach which gives very good results. It
is an exact approach for $C^k$-periodic solutions, ie for spatial derivatives up to order $k$. 
One of the reason to stop as some given $k$ is mainly practical: most often constraining $C^\infty$ regularity
for NNs is both useless and impossible (unless using very constrained NN architectures \cite{periodic_dong}). 
In the examples considered $k=2$ or
$k=3$ is in fact enough. Diminishing the regularity ($C^0$ or $C^1$) is not advised however, having visible
impacts on the class of minimizers obtained. The method relies on Hermite interpolation formula.
We give here the ansatz formula for $k=2$ in 1-D only with ${\cal D} = (0,1)$ for illustration.
\begin{equation}\label{perbc}
{\cal N}_{\rm per}(\tau,x) = 
$$
$${\cal C}_0 + {\cal C}_1 x + \frac12 {\cal C}_2 x^2
- ({\cal C}_1 + \frac12 {\cal C}_2) x^3 + (3 {\cal C}_1 + \frac12 {\cal C}_2) x^3(x-1) 
- 6 {\cal C}_1 x^3 (x-1)^2 + x^3 (x-1)^3 {\cal N}(\tau,x).
\end{equation}
Here ${\cal C}_{0,1,2}: \tau \mapsto {\cal C}_{0,1,2}(\tau)$ are three independent 
NNs with input dimension equals to $d=1$. For instance, one can easily 
check that ${\cal N}_{\rm per}(\tau,0) = 
{\cal N}_{\rm per}(\tau,1) = {\cal C}_0(\tau)$ for all $\tau$.
Such ansatz must be plugged in the cost functional whenever ${\cal N}_{\rm per}$ is needed and 
the three (small) additional NNs have
their own parameters to optimize with the other parameters during the full minimisation descent.

\subsection{Monte-Carlo sampling, stochastic and adiabatic gradient descent}\label{NN5}
We describe here rather standard procedures to solve 
the minimisation problem (\ref{generalC}) as well as useful tips.
\subsubsection{Monte-Carlo estimates}
The cost functional (\ref{pengeoa}) must be discretized. We follow the classical machine-learning
strategy which consists in using simple Monte-Carlo estimates. The basic idea is to  
draw a given number of random points in space from some ad-hoc probability distribution. 
The set of points obtained is often called {\it batch}.
There are important reasons for doing so. First, Monte-Carlo integrals do not suffer from
the curse of dimensionality so that one can easily handle higher dimensional integrals. 
Second, it is a very efficient way to deal with the strongly nonconvex behavior
of the NN optimisation problem. Due to that, the optimisation descent problem and NN parametrizations must be
considered as stochastic rather than deterministic. In our case, there is a particularity owning to the
nonquadratic structure of the geometrical action itself. We use here uniformally-distributed tensorial batches:
\begin{equation}\label{batch}
(\tau_i,x_j) \sim {\cal U}(0,1) \times {\cal U}({\cal D}), ~ i = 1,\cdots,N_\tau,~j = 1,\cdots, N_x.
\end{equation}
The integrals are then simple empirical means:
\begin{equation}\label{MC0}
\Int_0^1 ||f||_\chi~d\tau \approx \frac{\gamma_{{\cal D}}}{N_\tau} \sum_{i = 1}^{N_\tau} \left(
\frac{1}{N_x} \sum_{j=1}^{N_x} f(\tau_i,{\bf x}_j) (\chi f)(\tau_i,{\bf x}_j)
\right)^\frac12,
\end{equation}
where $\gamma_{\cal D}$ is a renormalisation factor (here the square root of the volume 
$\gamma_{\cal D} = \sqrt{|{\cal D}|}$). The auxiliary constraints (\ref{aux}) 
can be handled either using the same batches (\ref{batch}) or using spacetime samplings:
$$
(\tau_i,{\bf x}_i) \sim {\cal U}((0,1)\times {\cal D}),~ i = 1,\cdots N_{\tau x}.
$$
Other strategies with better precision might be considered as well, such as cubature formulas
and quasi-Monte-Carlo methods e.g. \cite{qmc}.

When monitoring the action values, there are two possible procedures. If the dimension $d$
is large, one can either compute the mean of the action estimate during the stochastic 
gradient descent (see below) over
a sufficiently large number of iterations, or use a single very large batch.
In moderate dimensions like in this work $d \leq 3$, it is often better to use 
Gaussian quadrature points or even Simpson rule on regular spacetime $d+1$-dimensional grids
and to perform a mean estimate over few iterations.
We emphasize that very precise action estimates must not be part of 
the minimisation algorithm itself, 
it is required in general at the very end. 
\subsubsection{Stochastic gradient descent}
Although widely used in machine-learning, we recall the very basics. The search for optimal NN parameters
$\T$ is performed through a stochastic gradient descent
\begin{equation}\label{sgd}
\T_{k+1} = \T_{k} - \eta \nabla_{\T} \overline{{\cal C}}_N|_{\T_k},
\end{equation}
where $\overline{\cal C}_N$ is the Monte-Carlo estimate of ${\cal C}$ (\ref{generalC}) 
(for instance using formulas (\ref{MC0}))
over a total of $N$ random batch points, $\eta$ is called 
the learning rate. The gradient of the discretized
cost functional $\overline{\cal C}_N$ is estimated using backpropagation algorithms, 
more generally, automatic differentiation (AD). These computations
must be fast and accurate. The AD algorithmic efficiency is indeed
one of the main reason for the success of deep learning methods.
The vector of parameters $\T$ is intrinsically random due to the use of batches
and (\ref{sgd}) might be interpretated as
a random dynamical system.  The larger the size of the batch $N$, the smaller the variance of the estimate $\T$.
More sophisticated first-order gradient descent are often used, in particular 
the so-called ADAM descent algorithm (see \cite{adam}). This is the default choice in this work.
The use of second-order deterministic quasi-Newton methods although superior, 
such as L-BFGS should be taken cautiously: they
are costly and should be used in deterministic context instead with careful initialisation procedure. 
The minimisation can be decomposed into two steps. First, a stochastic gradient descent (ADAM)
is performed until the networks become statistically stationary (if it is computationally affordable). 
Second, another descent 
can be performed to improve the precision by using a smaller learning rate, larger batches, or switching
to deterministic L-BFGS.

\subsubsection{Adiabatic descent}\label{adiad}
An important issue in machine learning is the initialisation of the NN parameters. By default, it is
random but requires a flux balance among different layers. A well-known procedure is the Xavier initialisation
\cite{xavier} and variants which are often tightened to the choice of the activation functions.
Very often, the convergence to relevant regions of the cost landscape is in fact very slow, or there
is no convergence at all. This situation happens either when the optimisation setup is highly penalized
(see \ref{NN4} with $\gamma_v,\gamma_w \gg 1$) 
or when the physical problem itself becomes ill-conditioned in some limit of interest, e.g. small diffusion
(see an example in \ref{ng2PDEs}), or extreme condition (see section \ref{burg1}).
A very powerful approach is to slowly perturb the problem from a situation where it is easy to solve
to a very stiff configuration. We call this an {\it adiabatic descent} since only one stochastic descent
is performed. We give the simplified algorithmic formulation for clarity. Let $\mu$ be an external parameter
(a physical one, some penalisation parameter/hyperparameter) and $\T$ the NN parameters to optimize.
\begin{itemize}
\item[(i)] Choose $\mu = \mu_0$ such that the problem is easy to solve, $\mu_{\rm end}$ a target problem
and $d\mu$ an adiabatic speed. Choose some incremental law $\mu \mapsto S(\mu,d\mu)$, e.g. a linear one 
		$S(\mu,d\mu) = \mu + d\mu$.
\item[(ii)] Perform a stochastic gradient descent for $\mu = \mu_0$ until satisfactory (statistical) convergence.
\item[(iii)] Do $k \to k + 1$ until $\mu_k = \mu_{\rm end}$
	$$\left\{
\begin{array}{llll}
	\mu_k    & = & S(\mu_{k-1},d\mu)) \\\\
\T_{k+1} & = & \T_{k} - \eta \nabla_{\T} \overline{{\cal C}}_N(\mu_k)|_{\T_k},
\end{array}\right.
$$
\end{itemize}
Some tuning is required here especially in the choice of the adiabatic speed $d\mu$ compared
to the inertia of the NN considered (typically the training rate): 
the change of $\mu$ must be small enough so that the
NN is able to relax to the new configuration. 
In practice, doing so enable the NN to better learn
the slow configuration changes as the weights and bias adjust more efficiently from a near optimal situation.
One must not try to perform a converged stochastic descent at each step since it would be too costly.
A remarkable aspect is that the NN size can be chosen rather small. This points towards the 
fact that the minimisation problem is most often the
issue rather than the expressivity of the network to capture stiffer solutions. A typical example is shown in
subsection \ref{gSDEs}.

The stiffer the configuration the smaller $d\mu$ must be. Therefore, one might often consider
nonlinear $S$ (e.g. a logarithmic change in the context of large deviations). Finally, we mention
that when $\mu$ is a physical parameter, it is important to perform an adiabatic search back and forth, 
ie $\mu_0 \to \mu_{\rm end} \to \mu_0$ in order to detect possible hysteresis phenomena.
Note also that this strategy is not very different, although more general,
from {\it scheduling} (and to some extent {\it early stopping}) 
which amounts to control the way the learning rate $\eta$ changes during the descent.

\subsection{Comparison with classical gMAM}\label{NN6}
We now discuss an important aspect: how does deep gMAM compare to the classical gMAM algorithm ?
The gMAM algorithm was first introduced in \cite{Hey_VdE} and used many times, e.g.
\cite{Hey_VdE_prl,tao,poppe}. Although, the minimum geometrical
action problem is exactly the same, the way it is solved is very different. For convenience denote
${\cal A}[u]$ the problem (\ref{geoa}) to minimize. Classical gMAM relies on finding the zeros of
the so-called Euler-Lagrange
(EL) equations which are formally $\delta {\cal A}[u] = \langle {\rm EL}[u],\delta u \rangle$.
For clarity, the FW action in its original form would give $\delta {\cal A} = \Int_0^T
\langle \dot u - F(u),\delta \dot u \rangle dt + 
\Int_0^T \langle \dot u - F(u), DF_u \delta u \rangle~dt$. It is already clear that in order
to obtain some explicit EL, one is forced to integrate by parts, as it is usually done.
Here it would be $EL = -\ddot u + DF_u \dot u + DF_u^\star (\dot u - F)$ where $DF^\star$ is the adjoint operator.
The classical gMAM approach therefore consists in solving $EL = 0$ using
the explicit form of the EL functional derivative. 
A simple way for doing this, is to consider a relaxation approach, by introducing a fictitious time 
$t$ and solving a preconditioned version:
\begin{equation}\label{EL}
\partial_t u = -\gamma_u EL[u].
\end{equation}
The discretization is in general in semi-implicit form. 
The term $\gamma_u = ||F(u)||/||\dot u||$ is the ad-hoc preconditioning quantity.
There is in addition a technical procedure that we do not detail here, see \cite{Hey_VdE,Hey_VdE_prl} 
where one must insure a constant arclength through interpolation every step or so. If not the 
(deterministic) gradient descent  becomes ill-conditioned. An important aspect
of this approach is the solution representation itself: $u(\tau,{\bf x})$ must be discretized
on a structured grid or using a Galerkin basis. 
\\\\
We can identify at least three important advantages of deep gMAM over the classical one. 
\begin{itemize}
\item First, the NN parametrization do not suffer from the curve of dimensionality with some outstanding expressivity of complex nonlinearities even in high dimension. 
Classical Galerkin decompositions are severely constrained by the curse of dimensionality. 
The minimisation problem (\ref{generalC}) can be handled in larger dimension $d$ without changing much
the NN size. Classical gMAM might involve very heavy optimisation problems since the size of the
		grid increases as $O(1/{\rm precision}^{d+1})$.
The number of NN parameters needed are typically known to behave polynomially with
the dimension instead (see e.g. \cite{beck2021deep}) and is the main reason for the great success of deep learning.

\item Second, at the level
of EL, one can anticipate that the integration by parts yields differential operators of order $2n$ whenever
$F$ contains differential operators of order $n$. Therefore, one expects that
the integration of (\ref{EL}) would then by tightened to more stringent CFL conditions. 
In the machine-learning context, EL equations
are found automatically (AD, backpropagation) in the neural network space, 
ie without having to integrate by part, see (\ref{sgd}).

\item Third, the stochastic aspect of the approach brings unexpected flexibility 
and efficiency. One can easily escape from local minima through stochastic fluctuations. 
In the deterministic gMAM, being trapped into a local minimum requires an external action.
The arclength condition is handled very easily at no cost and in fact 
the soft condition $\gamma_{\rm arc} \ll 1$ helps
the NN to explore better-adapted parametrizations.  Any additional constraints (see section \ref{NN4}) 
are straightforward to implement by simply adding more penalisation terms. 
This is not the case when dealing with an explicit EL approach, where it is often 
unclear how to constrain the problem in a simple way.

\item Another aspect which is worth to mention, is
that the differential operators are in fact exact approximations 
in the context of NN parametrization,
whereas classical discretizations relies on finite-order approximations schemes.
\end{itemize}
The main drawback of deep gMAM is the precision. Once classical gMAM converges, it has better precision than deep gMAM. The relative error typically encountered is $O(1\%)$.
This is in fact a much more general problem in deep learning, it is very difficult, although not
impossible to attain the same precision than classical methods. In general, costly procedures are needed to improve this. 
For instance, one can increase the NN capacity, fine tune the hyperparameters, or make use 
of quasi-Newton methods. It also includes the preconditioning of the cost functional in the neural network space: this is a well-known and very difficult problem in machine learning. 
It appears, however, that one is often more interested
in the generalisation behavior of NNs than the precision itself, especially when 
the dimension increases in which case classical approaches become hopeless anyway.
It is worth mentioning promising approaches to cope with the precision issue e.g. multi-scaled NNs
\cite{Cai1,Cai2} and domain decomposition \cite{DomDec}.
We summarize these remarks in the table below. 
\\
\begin{center}
{\footnotesize
\begin{tabular}{|c||c|c|}
\hline
& gMAM & Deep gMAM
\\ \hline \hline && \\
Discretisation & finite-diff, pseudo-spectral,... & NN parameter $\T$ (exact)
\\ && \\ \hline && \\
Descent vector & $-\frac{||F({\bf u})||}{\dot ||\dot {\bf u}||} \times$
\textcolor{black}{EL}, \textcolor{black}{Euler-Lagrange $\equiv \frac{\delta {\cal A}_g}{\delta u}$}
& $-\nabla_{\T} ({\cal A}_g + {\cal C}_{\rm arc} + \cdots)$
\\ && \\ \hline && \\
Descent algor. & {\bf deterministic}: relaxation, CG, L-BFGS & {\bf stochastic}: ADAM and variants
\\ && \\ \hline && \\
Arclength & hard constraint $||\dot {\bf u}|| = {\rm cst}$ (interpolation) & none (soft constraint)
\\ && \\ \hline && \\
Dim curse ($d \geq 3$) & Yes & No
\\ && \\ \hline && \\
 Non-convexity              & Moderate ($\sim$ continuous action) & Wild (NN par.)\\
\hline
Flexibility &  low & high \\
\hline 
Precision &  high & low \\
\hline
\end{tabular}}
\end{center}


\section{Examples: SDEs}\label{exSDE}
Although deep gMAM offers significative advantages in the context of PDEs $d \geq 1$, 
it also remains highly competitive in
the SDEs situation $d=0$ as well. The examples below are rather illustrative of the different scenarii
one can encounter in the context of PDEs. They should not be considered as a finality on their own.
We illustrate two important classes of systems (see also Appendix \ref{Julia}). The ones where the dynamics has a free-energy potential, the so-called
{\it gradient systems} and the others where there is no potential, ie 
the {\it non-gradient systems} where the time-reversal symmetry is broken.
In the first case, Arrhenius laws can be derived explicitly. In particular, it is possible
to show that the instanton trajectory must cross the saddle points (ie with Morse index one)
and orthogonal to the potential isosurfaces. Moreover, the instanton going from $a$ to $b$ has the same
geometrical support than the one going from $b$ to $a$ owning to the time-reversal symmetry
of the stochastic process. Only the value of the action may differ.

Nongradient systems offer much richer and unexplored behavior. In particular, instanton paths
do not need to cross saddle points in the separatrix. One can nevertheless recover an 
explicit Arrhenius law in the hypothesis of transversality \cite{tao}.
In fact, we give an example in subsection \ref{ngSDEs}
where the separatrix contains no saddle points.
The non-equilibrium trajectory therefore must go through a more complicated object playing the
role of the saddle.

\subsection{Gradient case: the (rugged-)M\"uller potential}\label{gSDEs}
The first example is the well-known M\"uller potential with two degrees of 
freedom (d.o.f.) \cite{Muller}. It is a gradient system ${\bf F} = -\nabla V$ with
$$
V = \sum_{i=1}^4 A_i {\rm exp}\left\{ {\alpha_i(x-X_i)^2} + \beta_i(x-X_i)(y-Y_i) + \gamma_i(y-Y_i)^2 \right\},
$$
where
$A = (-200,-100,-170,15)$, $\alpha = (-1,-1,-6.5,0.7)$, $\beta = (0,0,11,0.6)$,
$\gamma = (-10,-10,-6.5,0.7)$,$X = (1,0,-0.5,-1)$, $Y = (0,0.5,1.5,1)$. There are three minima
${\bf m}_1 = (-0.5582,1.4417)$, ${\bf m}_2 = (0.6235,0.0280)$ and ${\bf m}_3 = (-0.050,0.4667)$ and two saddles ${\bf s}_1 = 
(-0.8220,0.6243)$, ${\bf s}_2 = (0.2124,0.2929)$. We focus here on the transition ${\bf a} = {\bf m}_1 \to {\bf b} = {\bf m}_3$. We call ${\bf a} = {\bf m}_1$ and ${\bf b} = {\bf m}_3$. The Arrhenius law gives an action
$$
{\cal A}_{{\bf a} \to {\bf b}} = 2 ( V({\bf s}_1) - V({\bf a}) + V({\bf s}_2) - V({\bf m}_3)) \approx 229.1
$$
We penalize here the BV constraints $u(0) = {\bf a}$ and $u(1) = {\bf b}$, the other approach
described in subsection (\ref{NN1}) gives identical results. The details given below are just illustrative:
in fact any NN configuration would work. In other words, no fine tuning is required here. 

We now minimize the geometrical action using two NNs with 4 hidden layers having 10 neurons each, and one for each component $(x,y)$. One can use a single NN having a 2-dimensional vector input yielding the same result.
We impose here a large penalty on the boundary conditions with $\gamma_{\rm bv} = 1000$ and $N_\tau=512$. 
Much smaller $N_\tau = O(10)$ would also work.
Although not necessary, we also use a learning rate scheduling by decreasing it from $10^{-2}$ to $10^{-4}$ in a log linear way
over 20000 iterations. We recall that a single iteration is a random shuffle of $N$ batch points uniformally
distributed in $(0,1)$.
We obtain the mean value over the last ten iterations ${\cal C}_{{\bf a} \to {\bf b}} \approx 229.3$, ie a fairly close value to the Arrhenius estimate.
Note that closer values can be obtained if the batch size and the NN size is increased.
The instanton path which is a minimum energy path in this case compares very well 
with known results \cite{Muller,Mullervde}.
\begin{figure}[!h]
\centerline{
\includegraphics[width=9cm]{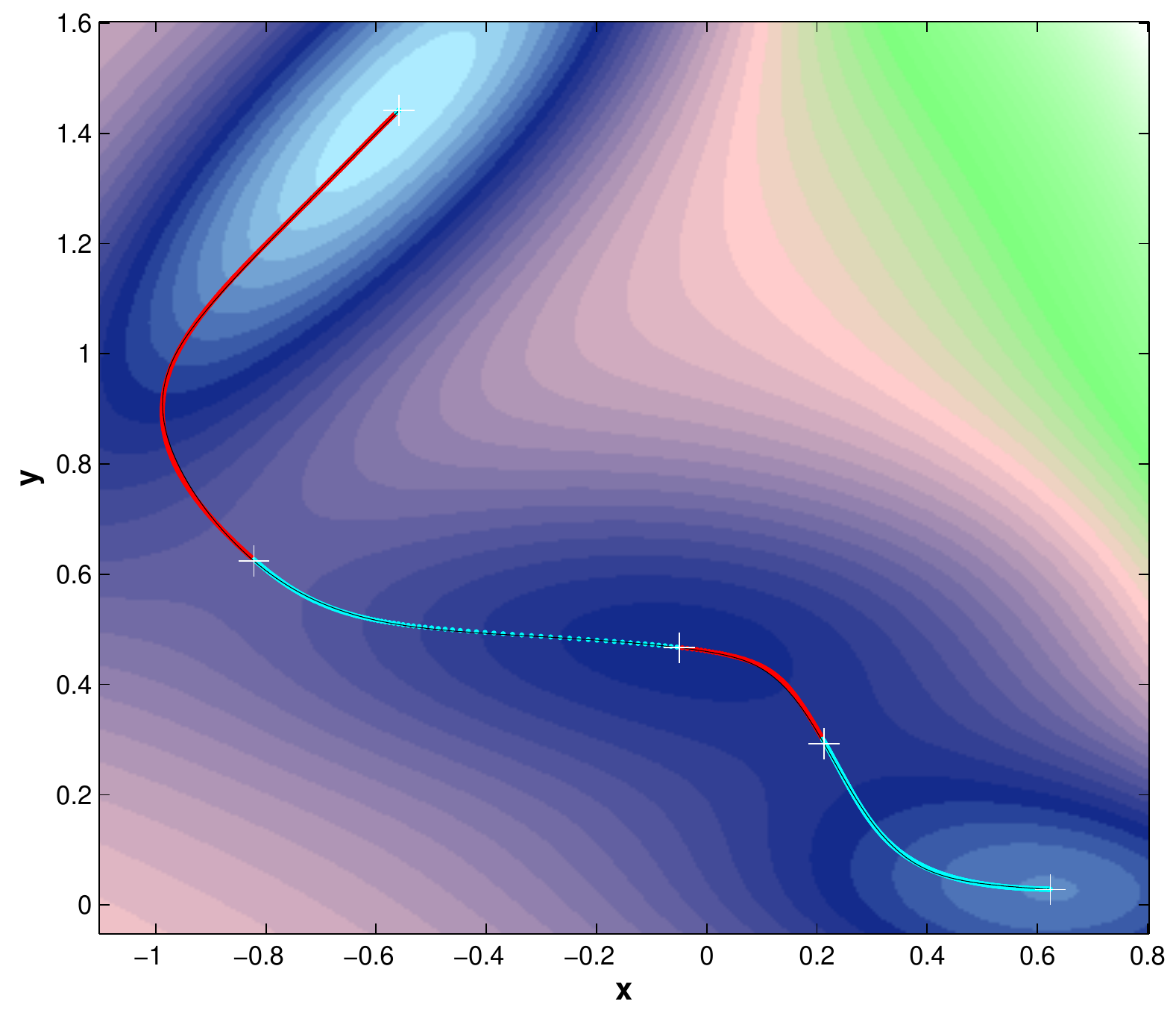}
}
\caption{Instanton ${\bf a} \to {\bf b}$ obtained  using a feedforward NN with 4 hidden layers having
10 neurons in each layers and swish activation function. Superimposed is the solution obtained from the classical
gMAM algorithm (black curve).
The red parts of the curve correspond to an the action above 0.2 (reactive path) and the complementary parts where the action is below
this threshold (relaxation path). The white crosses are the minima and saddles and 
the M\"uller potential is the colored background.}
\label{mul0}
\end{figure}

\begin{figure}[!h]
\centerline{
\includegraphics[width=17cm]{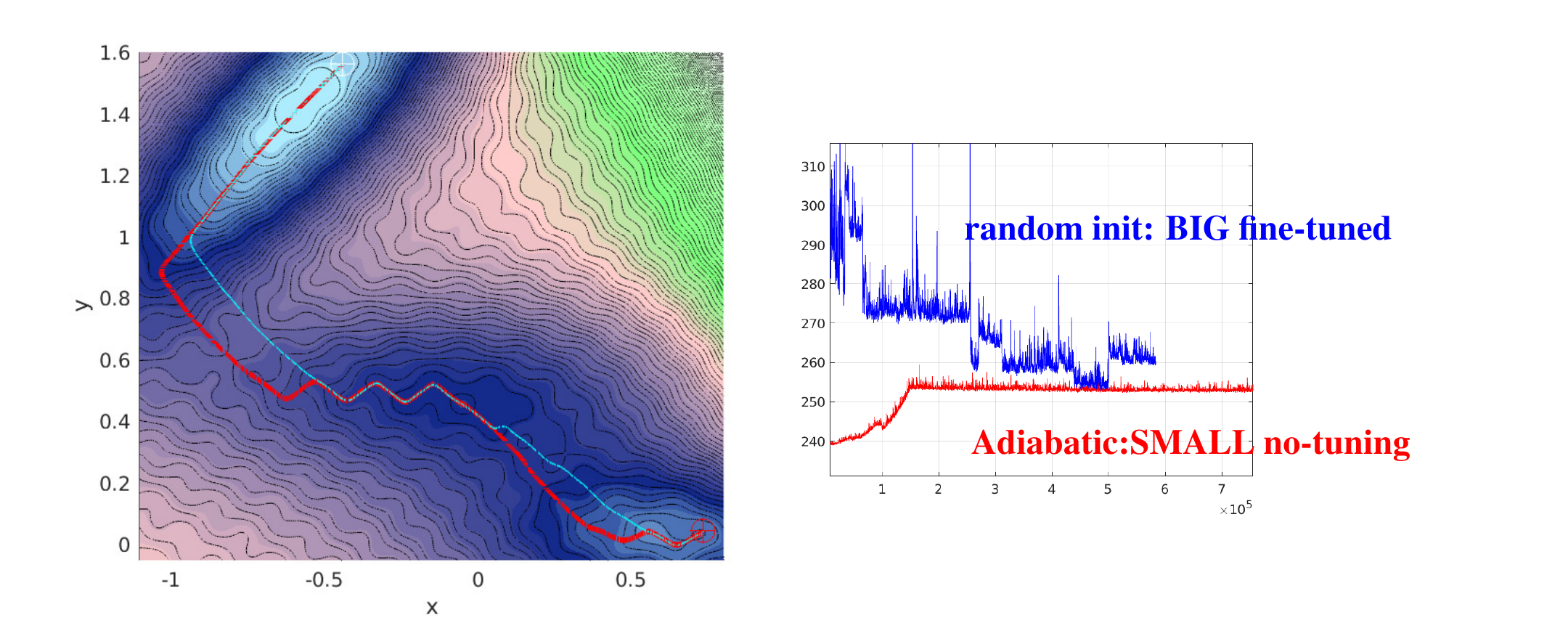}
}
\caption{Adiabatic method for $\rho$: in red we use a medium size NN with $8$ hidden layers with
10 neurons capacity and swish activations giving a total of 922 NN parameters and we slowly
change $\rho$ from $0$ to $3$ over $1.5 \cdot 10^5$ iterations 
(red curve right panel) and then froze $\rho = 3$. In blue, we purposely use a much bigger NN having 
20 hidden layers with skip connections giving a total of 11220 parameters and start from a random
NN initialisation (blue curve right panel). In this case $\rho$ is fixed $\rho = \rho_{\rm end} = 3$. 
In this example, we use formulation (\ref{ansatz}) and $N_\tau = 256$. 	}
\label{A_V_B}
\end{figure}

We now consider a more challenging situation by considering the so-called "Rugged M\"uller" version. One adds to the potential a high-frequency
perturbation:
$$
V_{\rm rugged} = V + \rho \sin 2 \pi n x \sin 2 \pi n y.
$$
We choose here $\rho = 3$. We are in fact able to handle larger values $\rho \leq 10$ in a similar way.
The potential landscape $V_{\rm rugged}$ has now many saddles: there are thus many local action mimimizers
and the total number of the action functional minima 
in fact explodes exponentially with the number of saddles of $V_{\rm rugged}$. Although
$d=0$ with only two degrees of freedom, the difficulty to capture the 
correct global minimizer is much
more challenging than gradient PDEs having much less saddles.
We illustrate the adiabatic descent strategy described in subsection \ref{adiad} in Fig. \ref{A_V_B}. 
The efficiency of the approach is rather striking, starting from random initial conditions turns
out to be very difficult whereas the adiabatic technique in cheap and easy. 
We believe that the red curve in Fig. \ref{A_V_B} is in fact the global energy minimizer,
although no known method is able to confirm it is actually the case. 
The classical gMAM method fails at giving lower-energy states than the ones obtained here.
Note that finding low-energy action curves for this example
is in fact physically irrelevant and other actions are 
required (E.Vanden Eijnden, private comm.). We therefore consider this example only as an
algorithmic illustration. 
 
\subsection{Nongradient case: hyperbolic limit cycle in the separatrix}\label{ngSDEs}
We now consider a striking example introduced in \cite{tao} where there is no saddle point in the
separatrix.
It is a 3 degrees of freedom (dof) nongradient system with
\begin{equation}\label{flow3}
\left\{
\begin{array}{llll}
F_1(x,y,z) & = & -(z+1)(z-2) \frac{x}{(x^4+y^4)^{1/4}} - x - y^3 \\
F_2(x,y,z) & = & -(z+1)(z-2) \frac{y}{(x^4+y^4)^{1/4}} + x^3 - y \\
F_3(x,y,z) & = & -(z+1)(z-2)z
\end{array} \right.
\end{equation}
This system has two stable fixed points ${\bf a} = (0,0,-1)$ and ${\bf b} = (0,0,+2)$ and a hyperbolic 
limit cycle $\Gamma$ on the separatrix at $z = 0$
satisfying $x^4+y^4 = 16$ with $\dot x = -y^3$ and $\dot y = x^3$. This limit cycle is stable on the separatrix $z=0$ and unstable in the $z$
direction. This system has the so-called transverse (or orthogonal) decomposition
\begin{equation}\label{transverse}
{\bf F} = -\nabla V + {\bf b},~\nabla V \cdot {\bf b} = 0,\forall {\bf x} \in \field{R}^3~
\end{equation}
with ${\bf b} = (F_1,F_2,0)$ and $V(x,y,z) = \frac{z^4}{4} - \frac{z^3}{3} - z^2$.  Due to that, it is possible to show
that the action value for the instanton going from ${\bf a}$ to ${\bf b}$ follows the classical Arrhenius formula (\cite{tao}, Theorem 1)
$$
{\cal A}_{\bf a \to b} = 2 \left( V({\bf x}_s) - V({\bf a}) \right).
$$
where ${\bf x}_s$ is any point belonging to the limit cycle $\Gamma$. In this 
case the potential depends on $z$ only
so that the minimum action is ${\cal C}_{\bf a \to b} = \frac{5}{6}$.
Note that the path $(0,0,-1+3s),~s \in (0,1)$ connecting directly ${\bf a}$ to ${\bf b}$ is
not a valid solution, although it has the minimal action value $5/6$, this path indeed coincides with the flow singularity $x=y=0$.

\begin{figure}[!h]
\centerline{
\includegraphics[width=7cm]{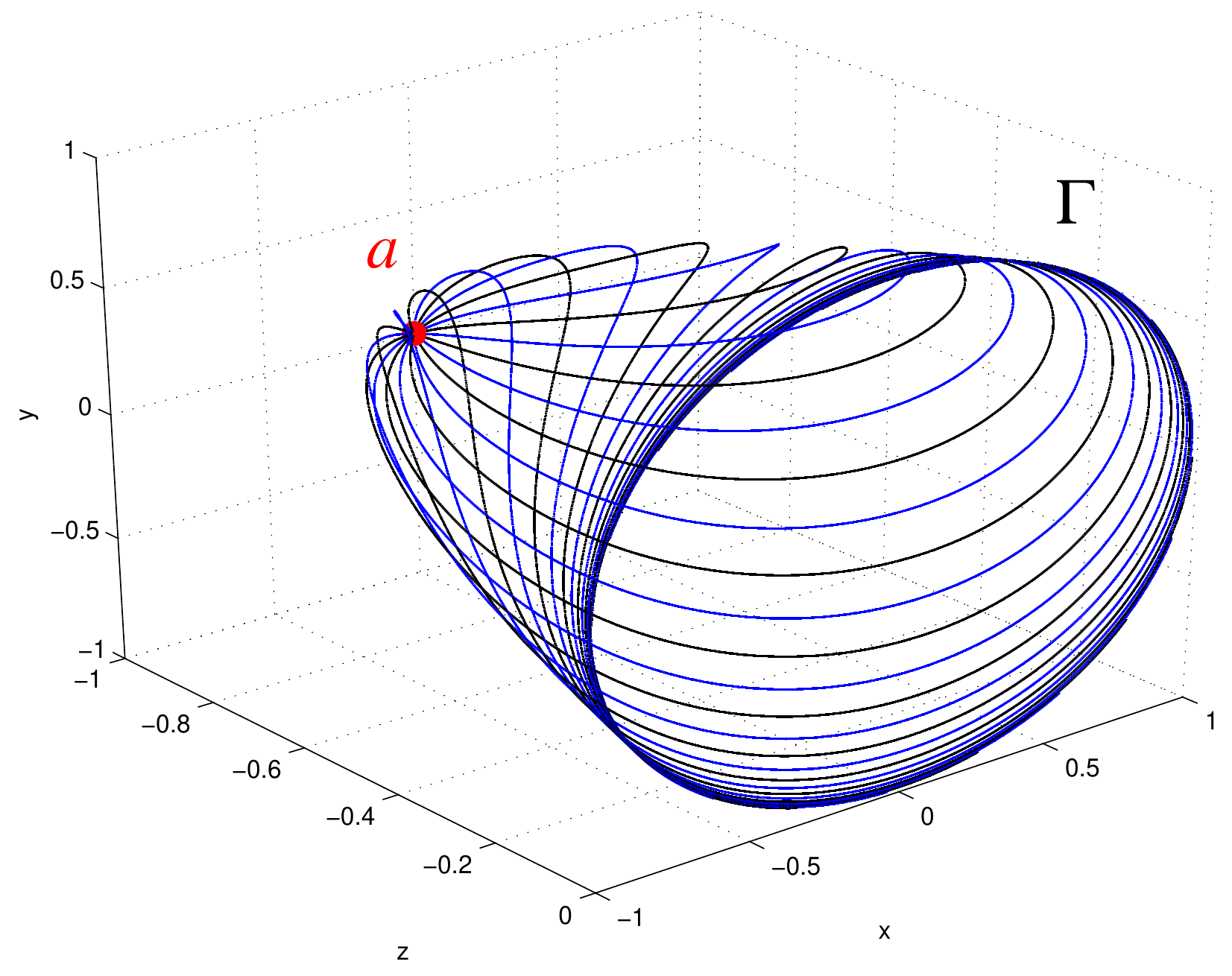}
}
\caption{There is a continuum of instantons connecting $a$ to the hyperbolic limit cycle $\Gamma$ on the separatrix ($z=0$),
all of them have the same action and are of
infinite length. This example is extracted from the system ${\bf F} = (1-z^2-r,1,z-z^3)$ 
in cylindrical coordinates $(r,\theta,z)$ (see \cite{tao}). 
Only two instantons are shown (blue and black curves) are $x(t) = r(t) \cos (t + \varphi), y(t) = r(t) \sin(t + \varphi)$ and $z(t)$ with
$r(0)=0,z(0) = -1$ and $r(\infty) = 1$,$z(\infty) = 0$ and $\varphi$ is arbitrary.}
\label{sketch}
\end{figure}

\begin{figure}[!h]
\centerline{
\includegraphics[width=6cm]{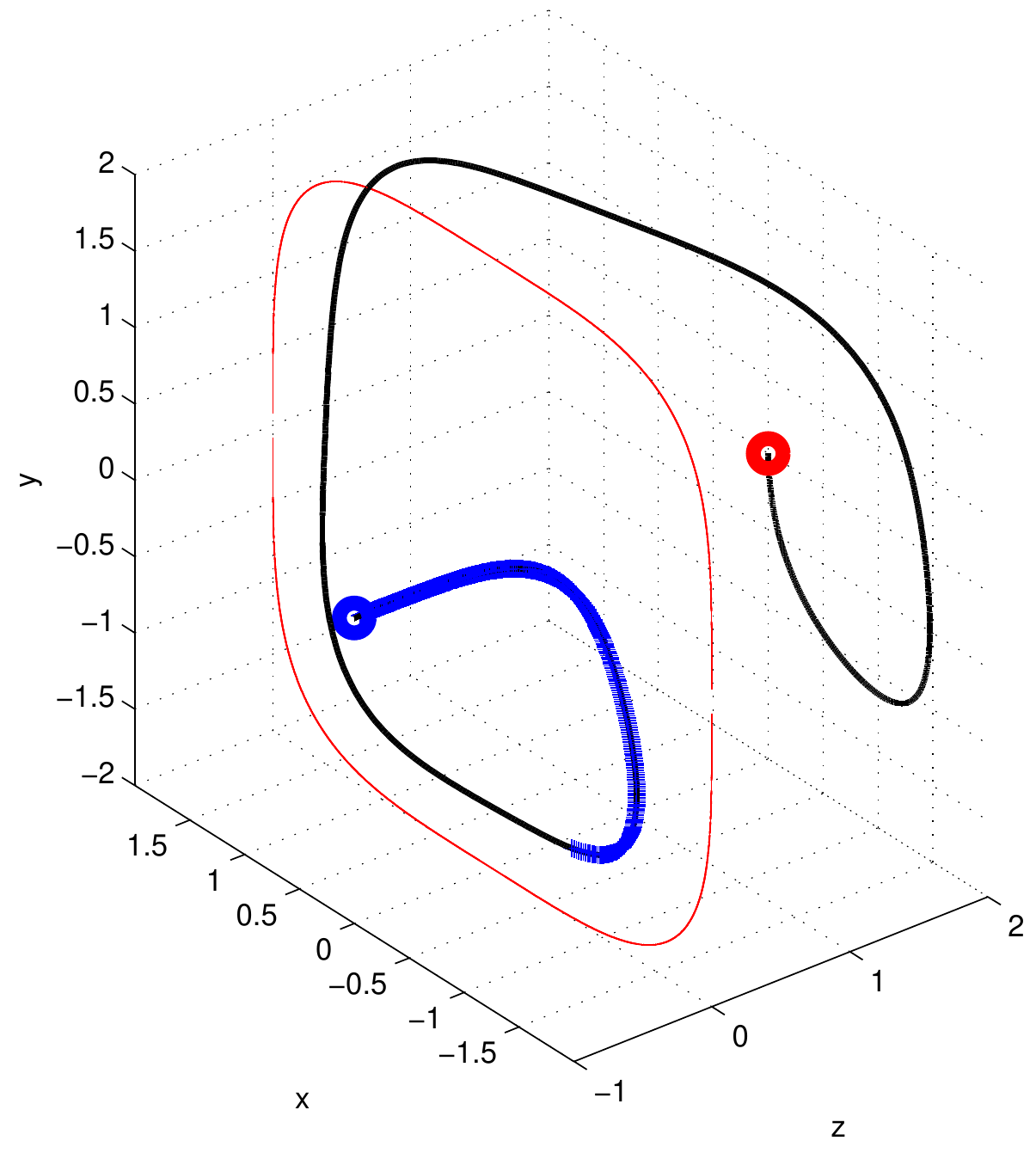}
\includegraphics[width=6cm]{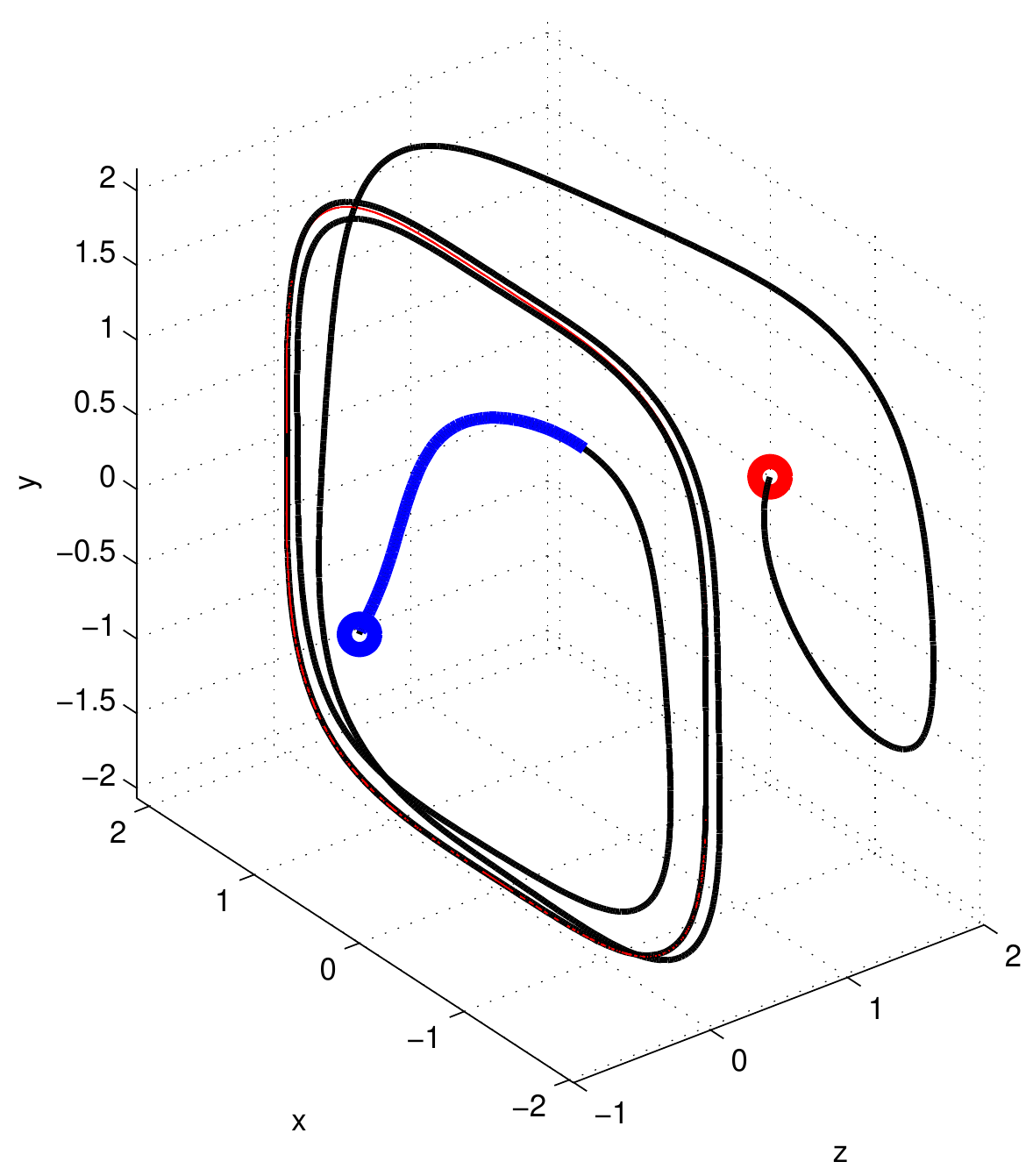}
}
\caption{Instantons connecting 
${\bf a}$ (blue dot) to ${\bf b}$ (red dot) for the flow (\ref{flow3}) using a NN with no arclength
condition (left panel) and an imposed arclength $L=50$ (right panel). The 
left panel gives an arclength $L \approx 17.7$. The hyperbolic limit cycle is shown in red 
in the $z=0$ separatrix hyperplane. The thick blue part of the curve is where the action is largest.
The total action is ${\cal A}_{\bf a \to b} \approx 0.838$ (left panel) and $\approx 0.837$ (right panel) 
(to be compared with $0.83333$).
NN size is $14$ hidden layers with 10 neurons for each layer using
swish activation functions. The descent algorithm is ADAM with learning parameter $\alpha = 10^{-3}$ 
and batch size $N=512$. The penalty
parameter $\gamma_L = 1$ for the right-panel experiment and $\gamma_{\rm bv} = 10$ for both.
}
\label{L50}
\end{figure}

It turns out that it takes an infinite time for the instanton to approach the separatrix. Due to 
the rotation, it winds around the limit cycle an infinite number of time. There are two consequences. 
First, the instanton length must be infinite.  Second, there are actually an infinite number of
instantons having the same action value which differ by a phase. It contrasts with 
the classical scenario where there is an unique instanton with finite length
going through a saddle point. A sketch of this situation is shown in Fig. \ref{sketch}
\\\\
We now show the results obtained by the deep gMAM algorithm in such a case. 
Figure \ref{L50} corresponds to two different experiments. The first one does not impose any condition on the arclength so that the NNs choose their
own parametrization. In the second experiment, we purposely impose a longer arclength for the solution 
(here $L=50$ versus $L\approx 17.7$ previously) by adding the constraint (\ref{instleng}).

We recover the fact (see Fig. \ref{L50}) that when we allow for longer arclength instanton, we obtain a solution winding around the hyperbolic
limit cycle $\Gamma$ like in Fig. \ref{sketch}.
The obtained values for the action is fairly close to the theoretical one
in all the experiments (see Fig. \ref{L50}). The action value between the two cases are very
close: the winding around the limit cycle contributes very little to the total action.
Note that in this case, no fine tuning of the hyperparameters is required and smaller networks would work
well.

\subsection{SDE quasi-potential: Maier-Stein system}\label{msqv}
We illustrate here as a byproduct of the method an efficient and simple way to compute
the quasi-potential (see definition (\ref{qv})). 
Consider an ODE having $N$ d.o.fs $\dot u = F(u)$, for simplicity we assume $\chi = Id$.
We want to compute (slices of) the quasi-potential $V$ in some $p$-dimensional
subspace $b \in {\cal D}_b \subset \field{R}^N$ and for $a$ fixed: 
$$
V: \field{R}^N \to \field{R}, b \mapsto V(a,b) = \inf_{u \in {\cal G}_{a,b}}
\Int_0^1 \left( ||\dot u|| ~||F(u)|| - \langle \dot u,F(u) \rangle \right)~d\tau,
$$
where ${\cal G}_{a,b} = \{ u(\tau) \in \field{R}^N, u(0) = a, u(1) = b \}$. 
The domain ${\cal D}_b$ can be anything. When $p=N$, one considers 
the full quasi-potential and when $p < N$ only lower-dimensional slices of $V$. 
Deep gMAM offers a very simple way to handle this problem by considering a larger network:
$$
{\cal U}(\tau,b): \field{R}^{p+1} \to \field{R}^N, ~p \leq N
$$
and minimizing the sum over $b$ of all geometrical actions:
\begin{equation}\label{allact}
{\cal C}[{\cal U}] = \Int_{{\cal D}_b} db \Int_0^1 \left( ||\dot {\cal U}|| ~||F({\cal U})|| 
- \langle \dot {\cal U},F({\cal U}) \rangle \right)~d\tau.
\end{equation}
This means that the neural network is encoding all the instantons at once and provides 
an estimate of the quasi-potential by summing over $\tau$, ie evaluating the geometrical
action at prescribed points once the minimisation has converged.
Classical approaches e.g. \cite{Cameron} rely on solving the stationary Hamilton-Jacobi PDE:
$2 \nabla V \cdot F + |\nabla V|^2 = 0, V(a) = 0$.
It is a notoriously difficult problem with multiple (viscosity) solutions and requires
careful numerical strategies. In our case, there is no particular treatment. 
It moreover offers the advantage to be a purely {\it local} approach: there is no constraints
on ${\cal D}_b$. It is only limited by the NN capacity to represent high-dimensional
ending conditions $b$.

In practice, it is better considering a BV ansatz (\ref{ansatz}) so that the procedure is 
transparent and does not need to monitor some BV penalisations. One needs to 
generate batches in the space $(0,1) \times {\cal D}_b$. In principle, one must 
build tensorial batches like in (\ref{batch}) of size $N_\tau \times N_b$. 
We show below that it is perfectly viable to consider $N_\tau = 1$ with 
consequence that the arclength condition (\ref{arc}) is by definition zero. We therefore remove 
the arclength condition in the example below. The example is a well-known testbed for nongradient
SDEs called the Maier-Stein system \cite{Maier}.
$$
F(u) =  (u_1 - u_1^3 -10 u_1 u_2^2, -(1+u_1^2)u_2).
$$
We fix $a = (-1,0)$ and $p=N$. The corresponding snippet Julia code is given in 
the Appendix \ref{Julia}. Note that the quasi-potential $V$ in nondifferentiable along the
$x$ axis which makes the convergence more difficult. In general, one has nonuniform convergence
of deep gMAM when the domain is large. When the domain is small, the NN imposes on the contrary
a spatially coherent global structure. It means that the various instantons tend
to share similar geometrical paths at the qualitative level. It is therefore a 
good practice to consider large enough NNs
in order to detect possible nondifferentiable structures correctly.

\begin{figure}[!h]
\centerline{
\includegraphics[width=11cm]{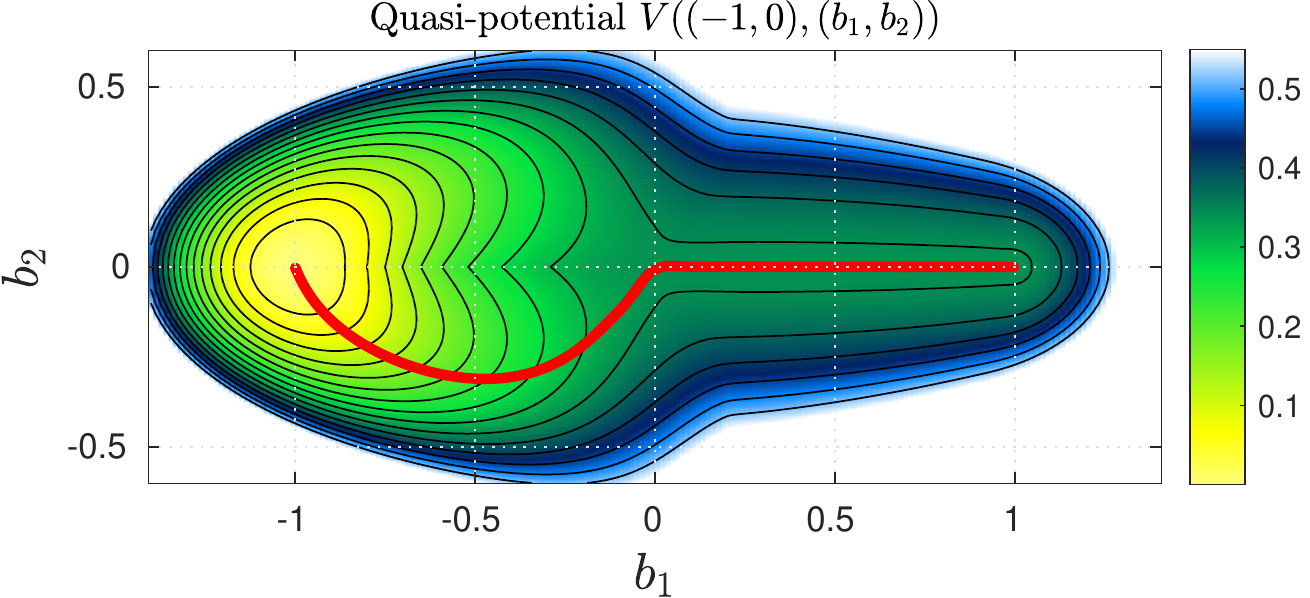}
}
\caption{Quasi-potential for the Maier-Stein system \cite{Maier} using the snippet code \ref{Julia}
with 10 hidden layers (15 neurons/layer, swish), with  $N_b = 1000$, $N_\tau = 1$.
Training rate is $10^{-3}$. The red curve is one of the 
two instantons connecting $(-1,0)$ to $(+1,0)$ (the
other is found by the symmetry $b_2 \to -b_2$).}
\label{QV}
\end{figure}


\section{Examples: SPDEs}\label{exSPDE}
The deep gMAM algorithm offers an even more interesting alternative in the PDE case where the trajectory now
lives in an infinite-dimensional functional space instead. 
A striking property is that one can use the same NNs used for SDE problems
to handle SPDEs problem: only the input dimension $d+1$ must be modified. It thus suggests that NNs
are very efficient to detect low-dimensional manifolds in very large-dimensional spaces.
We revisit some well-known examples of increasing difficulty
starting from $d=1$ (gradient PDE) to $d=3$ (nongradient PDE).

\subsection{Gradient case: Ginzburg-Landau}\label{gPDEs}

The Allen-Cahn-Ginzburg-Landau (ACGL) equation is very well documentated, starting from the seminal work of 
\cite{Jona}. It has been used many times to test ideas for computing non-equilibrium transitions due to 
its relative simplicity, e.g. in \cite{EVdE,Joran}.
The $d$-dimensional version with Dirichlet boundary conditions reads
\begin{equation}\label{acgl0}
\partial_t u  = F(u),~{\rm with}~F(u) \equiv \nu \Delta u - P'(u),~u = 0 \in \partial{\cal D}
\end{equation}
in the domain ${\cal D} = (0,1)^d$ and with
$$
P(u) = \frac14 (1-u^2)^2,~P'(u) = u^3-u.
$$
This PDE is in fact a gradient system which can be written as 
$\partial_t u = -\frac{\delta V}{\delta u}$, with the free energy potential
\begin{equation}\label{ACGLpot}
V(u) = \frac12 \Int_{\cal D} \left(  \nu |\nabla u|^2 + 2P(u) \right)~d{\bf x}.
\end{equation}
We use here an approximated formula of the stationary solutions, namely we use
$$
a({\bf x}) = \prod_{l = 1}^d \tanh \frac{x_l}{\sqrt{2\nu}} \tanh \frac{1-x_l}{\sqrt{2\nu}},~{\rm and}~ b({\bf x}) = -a({\bf x}).
$$
This approximation is in fact very close to the true solution up to exponentially small correction terms.
We want to compute the most probable transitions between the two metastable states $a$ and $b$. 
This problem has several group symmetries: the symmetric group ${\bf x} \mapsto \sigma({\bf x})$ where $\sigma$ is a permutation, the reflection
group ${\bf x} \mapsto (-1)^\sigma {\bf x}$ or the transformation mapping $u \to -u$.
We thus expect that any solution is in fact associated with several (at least 2$d!$ possibly identical) states and that many
symmetry-breaking bifurcations can occur. Typically, the trivial state $u_{s,0} = 0$ is invariant by all symmetry groups,
and the states $a,b$ are invariant by all the permutations and reflections.
Depending on the value of the diffusion parameter $\nu$, the transitions $a \to b$ can differ. For instance, a first symmetry-breaking
bifurcation occurs at a critical value $\nu^\star$ for which a new family of critical
points appears. In dimension $d=1$, there are two of them which
are antisymmetric w.r.t. the axis $x=\frac12$. They behave as $u_{s,1}  \sim \pm \tanh (x-\frac12)/\sqrt{2\epsilon}$ near $x = \frac12$.
In $d$ dimension, we expect $2 d$ saddle states each corresponding to an antisymmetric solution w.r.t. to the $d$ hyperplane $x_l = \frac12$.
We omit the exact discussion on these group symmetries and focus first on $d=1$.
It can be shown that a cascade of bifurcations occurs at the critical values
$$
\nu_n = \frac{1}{(n+1)^2 \pi^2},
$$
where pair of states $\pm u_{s,n}$ appear and correspond to fronts having $n$ zero in the interval $]0,1[$.
Since (\ref{acgl0}) has a gradient structure, one can use the Arrhenius formula:
\begin{equation}\label{Arr_ACGL}
{\cal A}_{a \to b}^{(n)} = 2 (V(u_{s,n})-V(a)).
\end{equation}
Very precise estimates can be obtained provided $\nu$ is small enough. For $n=1$, $V(u_{s,1}) = V(0) = \frac14$. We can estimate
also $V(a) \sim \frac{2 \sqrt{2}}{3} \sqrt{\nu}$ using the fact that the solutions are $\pm 1$ with boundary-layer corrections. An easy computation
gives
\begin{equation}\label{arrflip}
\mbox{flip transition:~~}{\cal A}_{a \to b}^{(0)} \sim \frac12 - \frac{4}{3} \sqrt{2\nu} 
\end{equation}
The expression for $n > 1$ are different, one can remark that both $a$ and the
critical points have the same behavior at the boundaries so that their contribution
cancel out. It only remains to compute the contributions of the zeros only which are $2n$ times those from the boundaries.
It comes
\begin{equation}\label{arrfront}
\mbox{$n$ front transition:~~} {\cal A}_{a \to b}^{(n)} \sim \frac{4}{3} n \sqrt{2\nu},~n \geq 1,~\nu \leq \nu_n.
\end{equation}
The consequence is that the global minimizer can be easily distinguished, if $\nu > \frac{1}{4 \pi^2}$, it is the flip solution and if
$\nu < \frac{1}{4 \pi^2}$ it is the $n=1$ front transition. All other solutions are local minima having larger action.

\begin{figure}[!h]
\centerline{
\includegraphics[width=5cm]{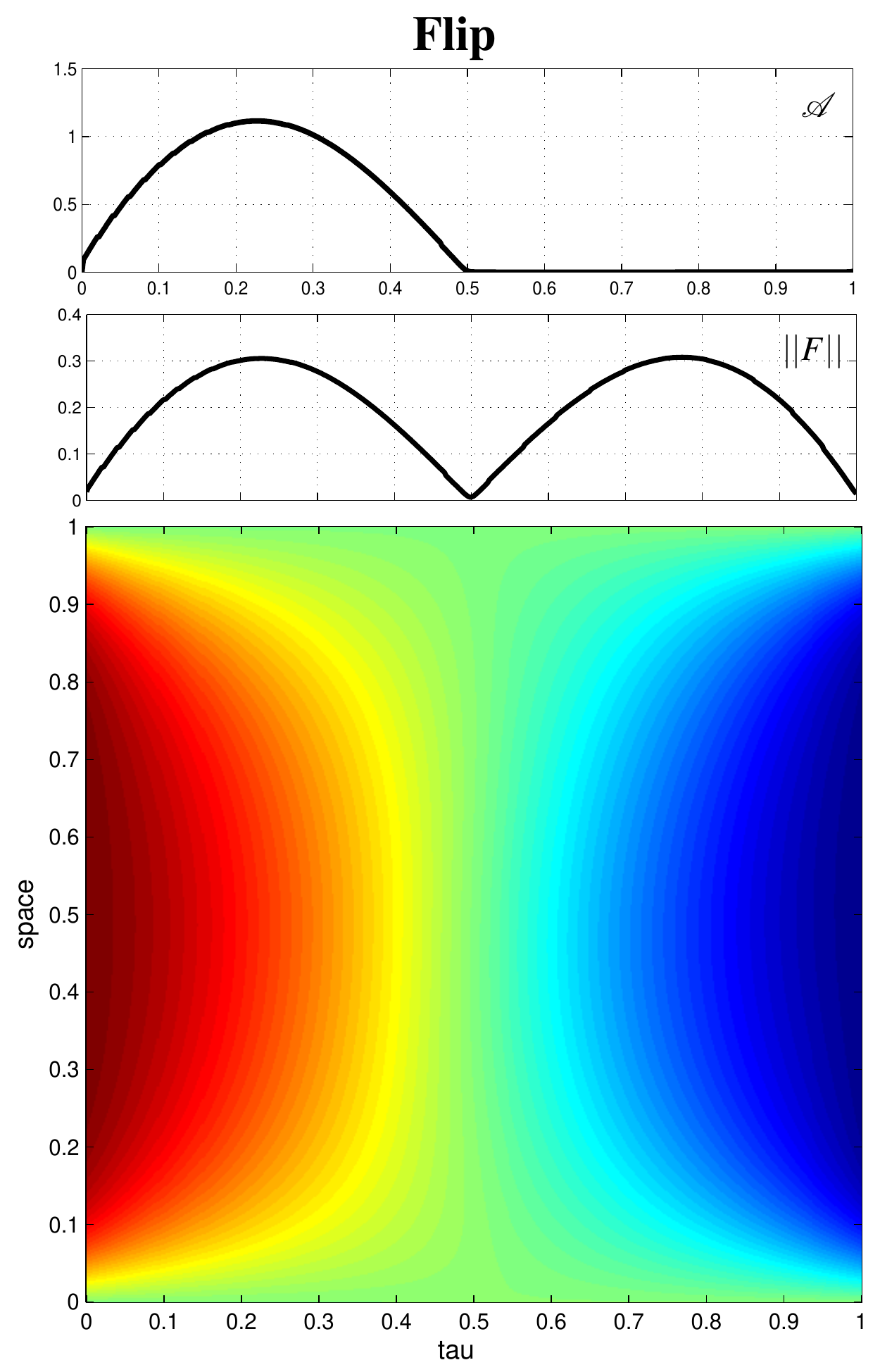}
\includegraphics[width=5cm]{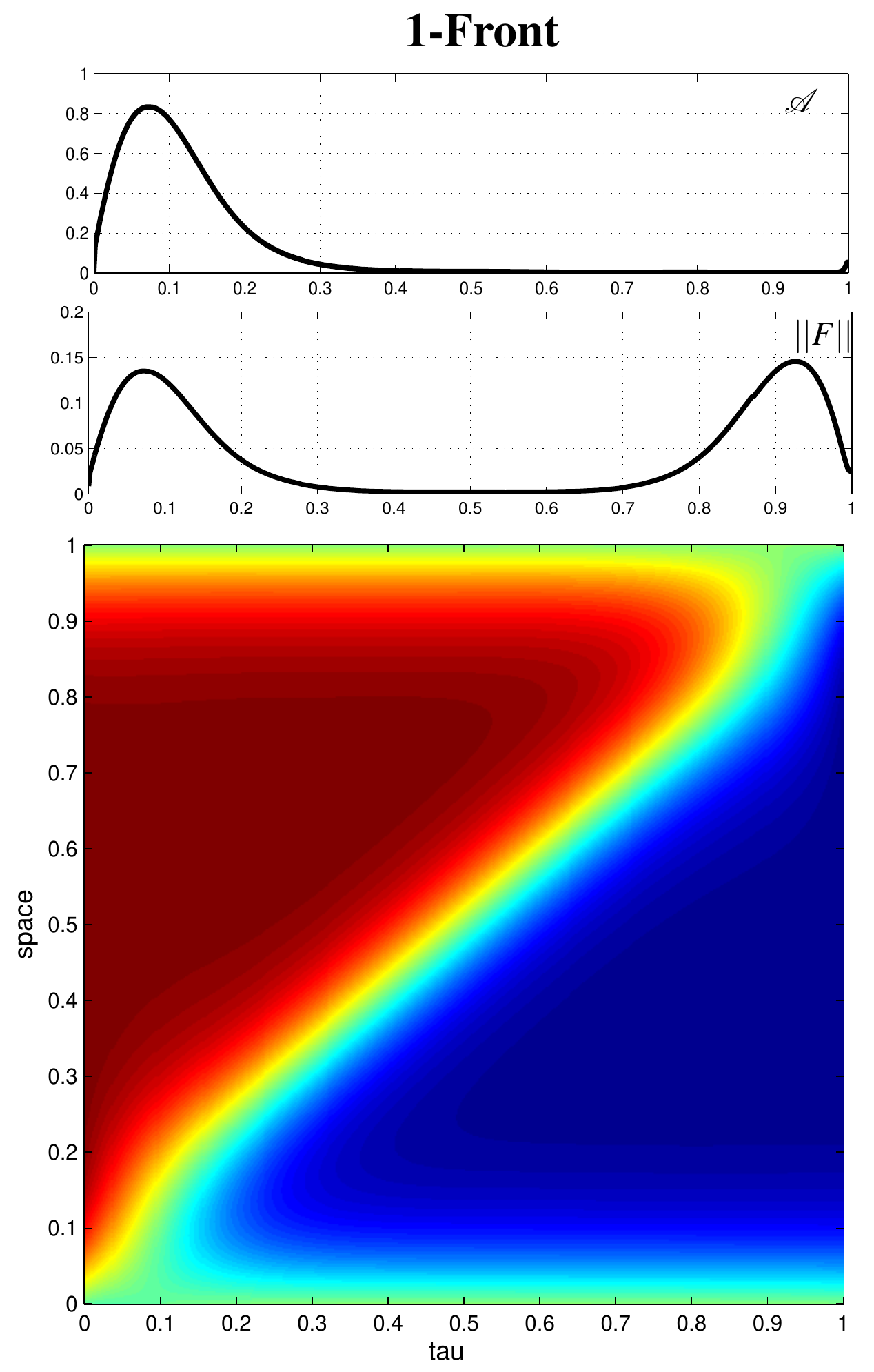}
}
\caption{Two transitions obtained using 5 hidden layers with 10 neurons each with swish activations for $\nu = 5\cdot 10^{-3}$: the flip transition (left) and the 1-front transition (right).
The mean action obtained over the last 100 iterations is ${\cal A}_{a \to b}^{(0)} \approx 0.36$ (left) and $A_{a \to b}^{(1)} \approx 0.13$ (right)
to be compared with the theoretical estimates $0.366$ and $0.133$ resp. The value of the action and the spatial norm $||F||$ are shown as a function
of the normalized arclength (ie computed only for visualisation) in the panels above.
The penalty parameters are $\gamma_{\rm bv} = \gamma_{\rm bcs} = 10$, and $\gamma_{\rm arc} = 0.1$.
The batch size is $N_\tau \times N_x = 16 \times 64$ and ADAM learning rate $\alpha = 10^{-3}$.
The two solutions have different total length: $L_{\rm flip} \approx 1.8$ and $L_{\rm 1-front} \approx 3.1$.
}
\label{GLINST}
\end{figure}

We now test the deep gMAM algorithm.
The results are shown in Fig. \ref{GLINST} and are in close agreement with the theoretical ones.
Only the 1-front solution is the true global minimum having smaller action.
We observe the typical reactive/relaxation parts for
both solutions: positive action before reaching the separatrix and zero-action when it relaxes to the target solution $b$.
Remarkably, we observe that for the front transition, the region around the saddle is very flat, confirming the analysis done in \cite{Joran}: for
finite-amplitude noise one indeed observes random-walk diffusion due to the landscape flatness.

\subsection{Nongradient case: $d=1$ Sheared Ginzburg-Landau}\label{ng1PDEs}
We now consider an example very similar to the previous one, except for two aspects: 1) it has 
an additional transport term $\partial_x u$
which makes it a nongradient version of the 1-D ACGL 2) we consider periodic boundary conditions. 
It is also taken from \cite{tao} and it corresponds to the same
scenario than in \ref{ngSDEs} but in the context of PDE: there is an hyperbolic limit cycle 
inside the separatrix so that transitions can
go through it rather than going through a saddle point. The PDE reads
\begin{equation}\label{acgl1}
\partial_t u = F(u),~ F(u) = \nu \partial_{xx}^2 u + u-u^3 + 0.1 \partial_x u,
\end{equation}
in the periodic domain ${\cal D} = (0,1)$. It appears that this equation 
also satisfies the transverse decomposition (\ref{transverse}) with the
same potential (\ref{ACGLpot}) than for the ACGL equation. The orthogonal term $b$ 
in (\ref{transverse}) is simply $b(u) = 0.1 \partial_x u$.
Any fixed points of the ACGL equation (ie with $c=0$) are in fact limit 
cycles of (\ref{acgl1}), since $u(x,t) = v(x+0.1 t)$ satisfies
$\partial_t u = F(u)$ (\cite{tao}). Due to that, one can obtain precise theoretical estimates of 
the possible transitions by using the same Arrhenius law (\ref{Arr_ACGL}) where $u_{s,n}$ are 
now periodic solutions, ie they satisfy
$$
\nu \partial_{xx}^2 u + u-u^3 = 0,~u(x+1) = u(x).
$$
It turns out that the solutions are the same than before except that the solutions with $n$ 
odd must be removed since they
cannot fulfill the periodic condition.
They can be studied in more details by interpretating space as time and 
analyzing the nonlinear oscillator $\nu \ddot u = u^3-u$ with the constraint
$u(t+1) = u(t)$ (see e.g. a detailed explanation in \cite{tao}). 
We can therefore use the same formula (\ref{arrfront}) but for $n=2p,~p \geq 1$.
In particular the mode with smallest action is now the mode with $n=2$.

In summary, we have a classical saddle-point transition going through the trivial saddle $u_{s,0} = 0$ 
and the transitions going through hyperbolic limit cycles:
\begin{equation}\label{arrshear}
\mbox{flip transition}: {\cal A}_{a \to b}^{(0)} = \frac12,~\mbox{LC $2p$-transition}: {\cal A}_{a \to b}^{(p)}
\sim \frac{8}{3} p  \sqrt{2\nu}.
\end{equation}
Note that for the flip transition, the solution is in fact explicit, it takes the 
form $u(s,x) = -1 + 2s,~s \in (0,1)$ and it is easy to
check that the geometrical action takes the value $\frac12$ in this case.

\begin{figure}[!h]
\centerline{
\includegraphics[width=5cm]{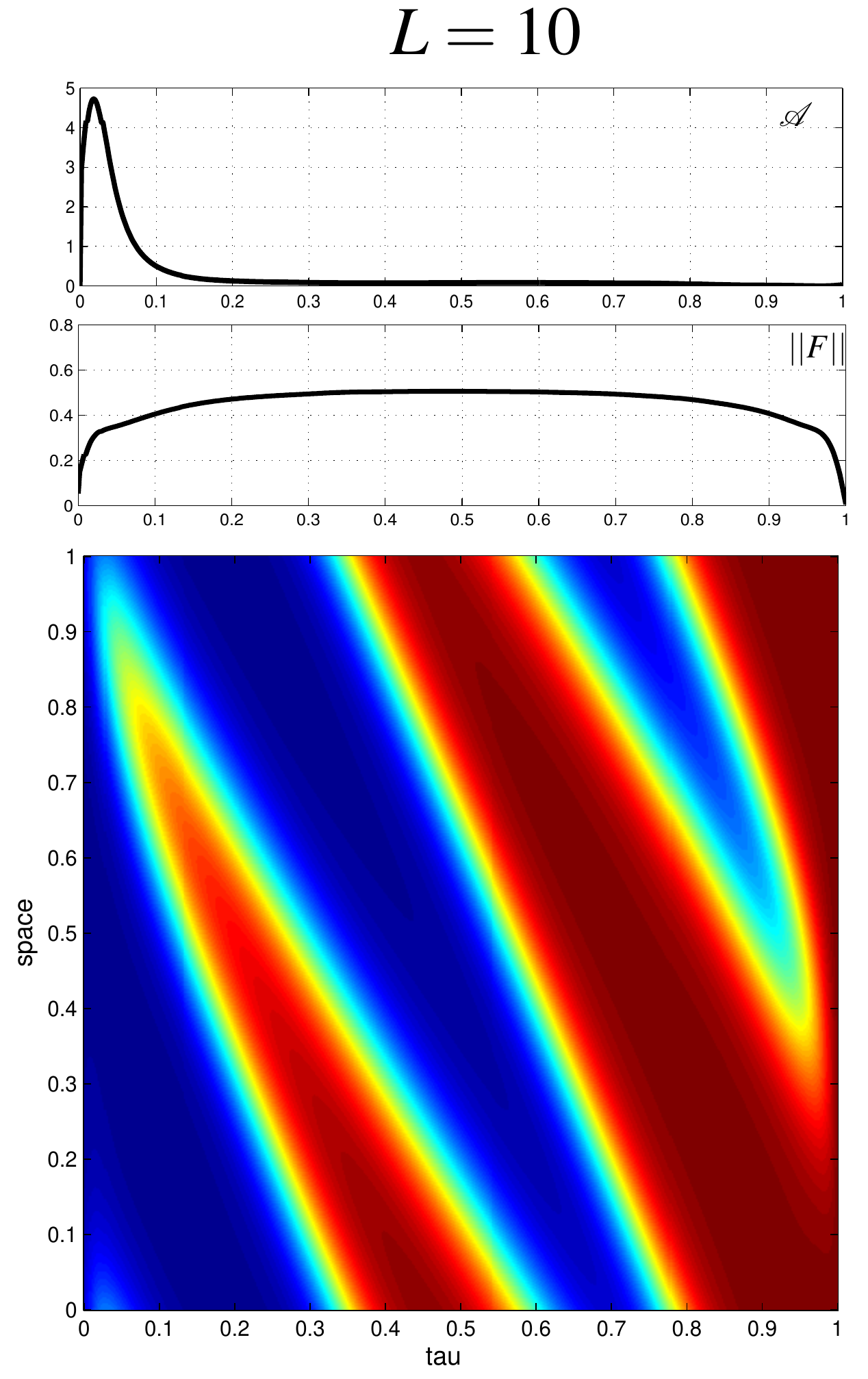}
\includegraphics[width=5cm]{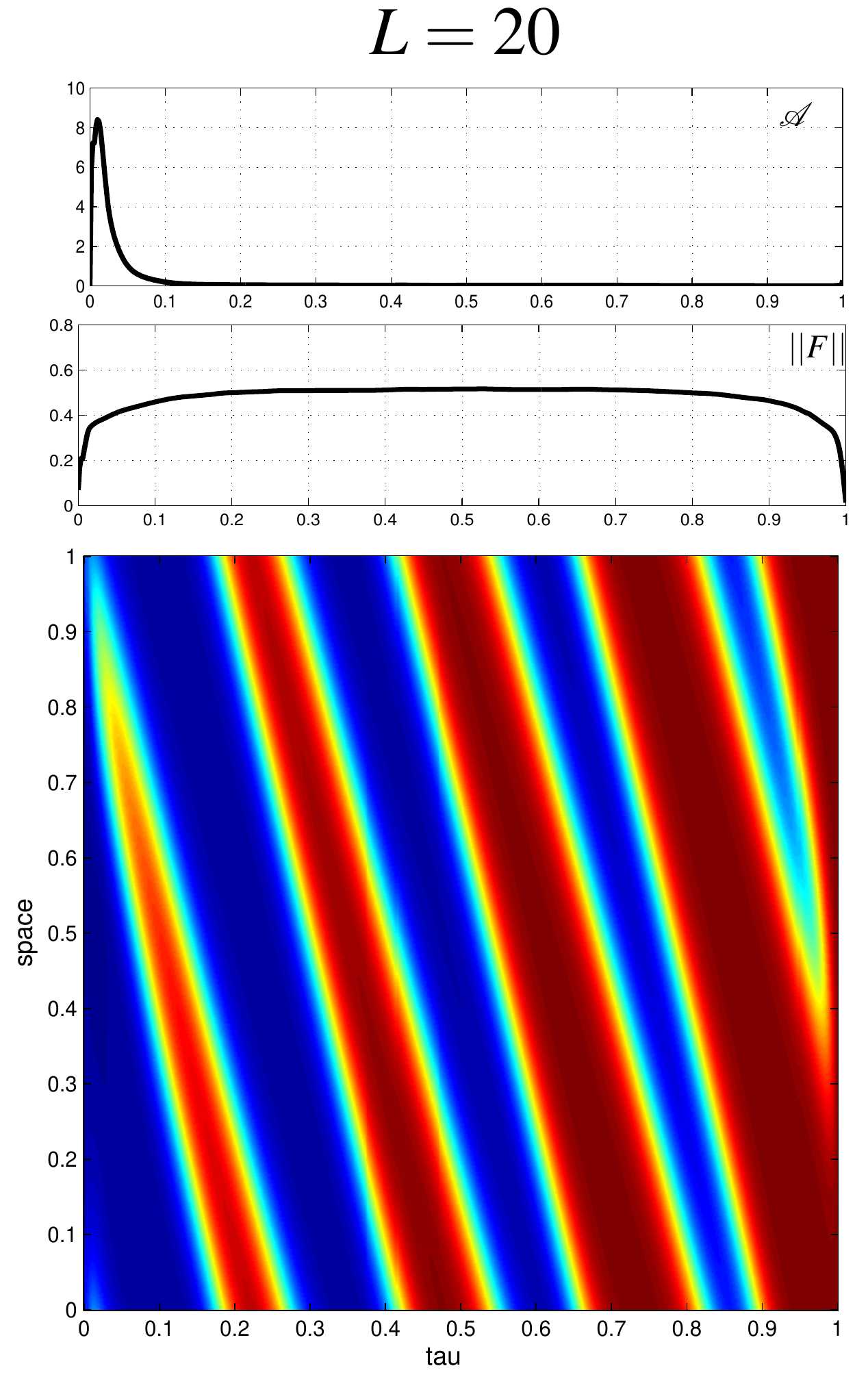}
}
\caption{Instantons of different total length $L$ and going through the $p=1$ hyperbolic limit cycle (see text).
The NN has 7 hidden layers with 10 neurons/layer with swish activations for $\nu = 5\cdot 10^{-3}$.
The value of the action and the spatial norm $||F||$ are shown as a function
of the normalized arclength in the panels above. Action values are ${\cal A}_{a \to b} \approx 0.32$ (left), and
${\cal A}_{a \to b} \approx 0.28$ (right).
}
\label{SGLINST}
\end{figure}

We now describe the numerical results. They are obtained for $\nu = 5 \cdot 10^{-3}$.
First of all, the trivial flip transition is easily obtained (not shown) and show the same type of behavior than in
Fig. \ref{GLINST} (left) although now the solution is periodic in space. We are interested in the global minimizer instead whose action is in this case $8/3 \sqrt{2\nu} \approx 0.266$ and from the discussion above, there is 
in fact a continuum of instantons of
infinite length going through an hyperbolic limit cycle, like in Fig. \ref{L50}.
We thus conduct two different experiments where we impose an instanton length $L = 10$ and $L=20$. 
The results shown in Fig. \ref{SGLINST} strongly
agree with the theoretical predictions. For instance, the norm $||F||$ does not vanish and is a very good indicator of the
type of transition. This can be used for instance to decide whether one must impose some total length 
condition or not, since the consequence of such a behavior is the nonuniqueness of the 
action global minimizers. The larger $L$ the closer the solution is
to the actual infinite-length instanton. We find that the short length solution overestimates 
the (mean) action value (Fig. \ref{SGLINST}, left),
whereas the longer one gives a smaller value more consistent with the theoretical value 
(Fig. \ref{SGLINST}, right). 

\subsection{Nongradient case: $d=2$ Sheared Ginzburg-Landau}\label{ng2PDEs}
\begin{figure}[!h]
\centerline{
\includegraphics[width=9cm]{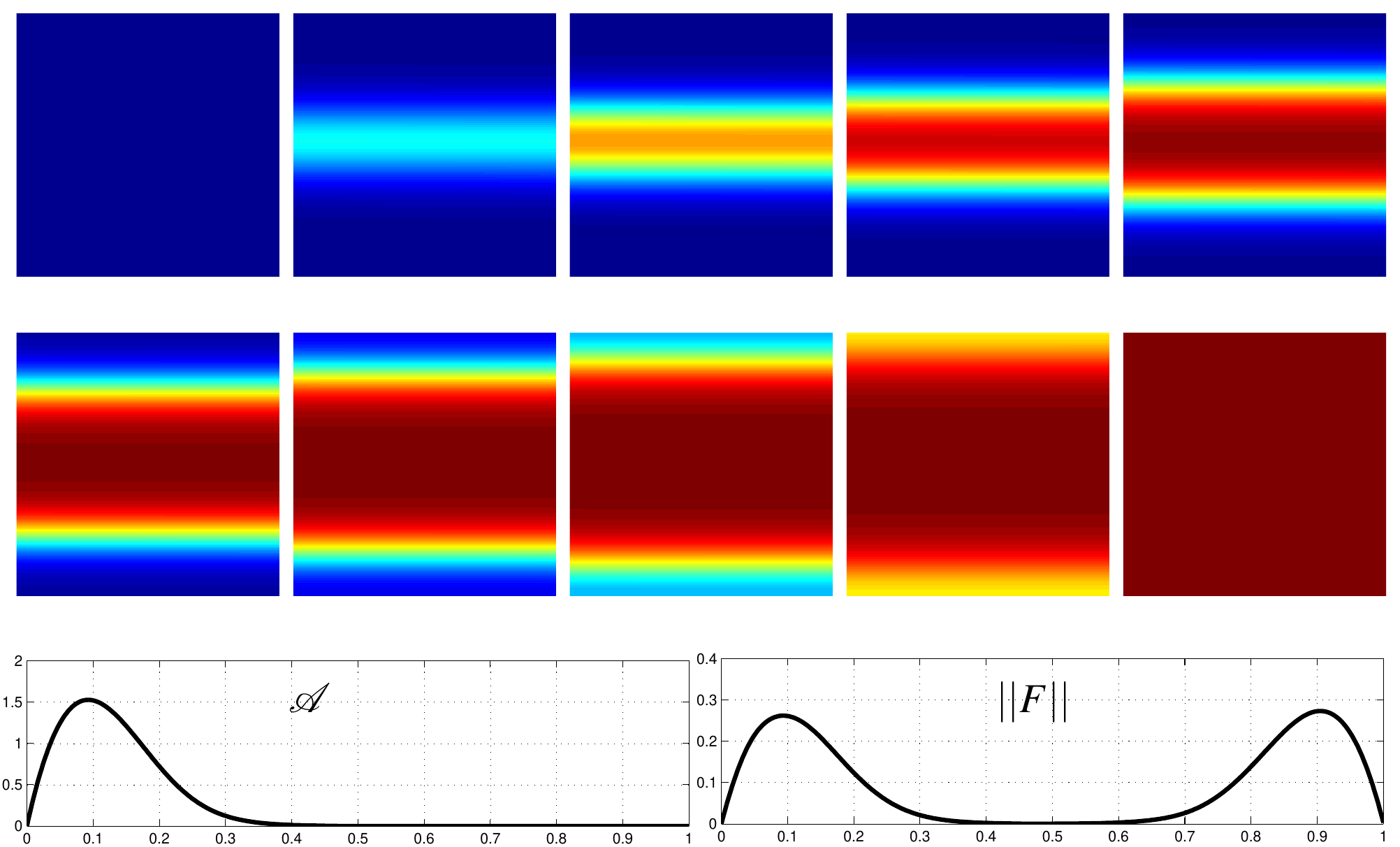}
}
\caption{A local $x$-invariant minimizer (nucleation) for $\nu = 5 \cdot 10^{-3}$ and $c=0.1$ for Eq.(\ref{2dshear}). 
Ten snapshots at equally distance are shown with color ranging from -1 (blue) to +1 (red). The 
NN has 7 hidden layers with 10 neurons with swish activations.
The value of the action and the spatial norm $||F||$ are shown as a function
of the normalized arclength in the panels below. The saddle corresponds approximatively to the panel 
(2,1) in the snapshot sequence. The
total length is $L \approx 2.9$ and the action ${\cal A}_{a \to b} \approx 0.269$ 
(to be compared to 0.266, see formula (\ref{arrshear})).
}
\label{ZON}
\end{figure}

This problem corresponds to a 2-D sheared version of the Ginzburg-Landau equation originally studied in 
\cite{OJK}. Action minimizers have already been investigated in \cite{Hey_VdE_prl} and \cite{tao} using 
the classical gMAM algorithm and can serve as useful benchmarks.
This example gathers all the difficulties together but now in spatial dimension $d=2$: it
is a nongradient PDE having both hyperbolic limit cycles and saddle fixed points
in the separatrix. It is moreover believed (see \cite{tao}) that this PDE cannot has the transverse decomposition  (\ref{transverse}). Indeed,
an example can be found where
two different instantons having different values of the action cross the same saddle fixed point, 
therefore contradicting Arrhenius formula
(\ref{Arr_ACGL}). The equation reads
\begin{equation}\label{2dshear}
\partial_t u = F(u),~F(u) = \nu \Delta u + u - u^3 + 0.1 \sin 2\pi y \partial_x u,
\end{equation}
with $a = -1,~b = +1$ as before, in the periodic domain ${\cal D} = (0,1)^2$.
The set of action minimizers contains  in particular the solutions which do not depend on $x$, 
ie the ACGL 1-D solutions.
A global minimizer candidate is therefore the one corresponding to (\ref{arrshear}) with 
$p=1$ and going through the corresponding
saddle-fixed point. It is shown in Fig. \ref{ZON}.

\begin{figure}[!h]
\centerline{
\includegraphics[width=9cm]{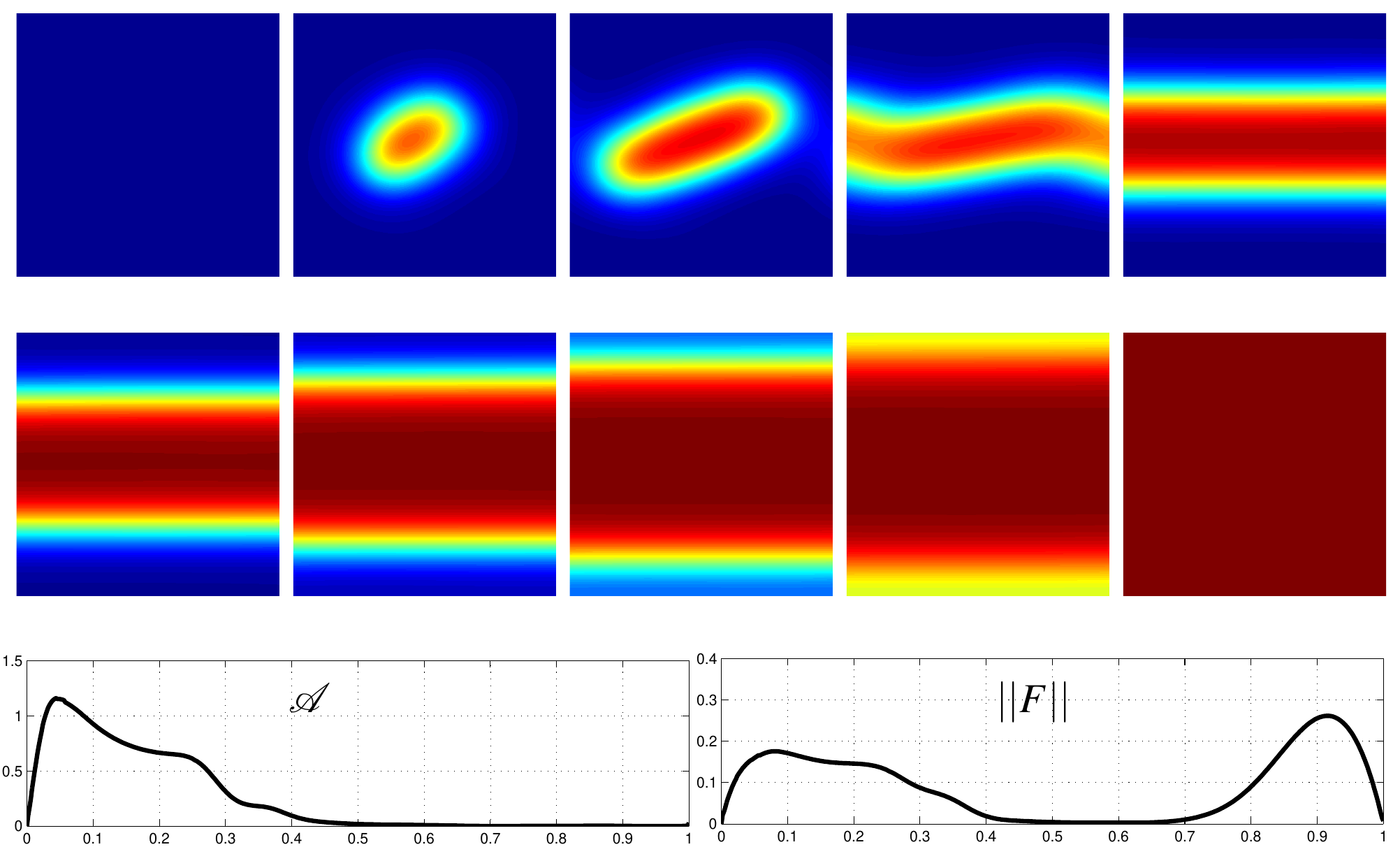}
}
\caption{Same as in Fig. \ref{ZON}. The nucleation now breaks the $x$-invariance and is 
	believed to be the instanton global minimizer (to be compared with \cite{Hey_VdE_prl}).
The total length is $L \approx 3.3$ and the action ${\cal A}_{a \to b} \approx 0.247$.
}
\label{SFI}
\end{figure}

\begin{figure}[!h]
\centerline{
\includegraphics[width=15cm]{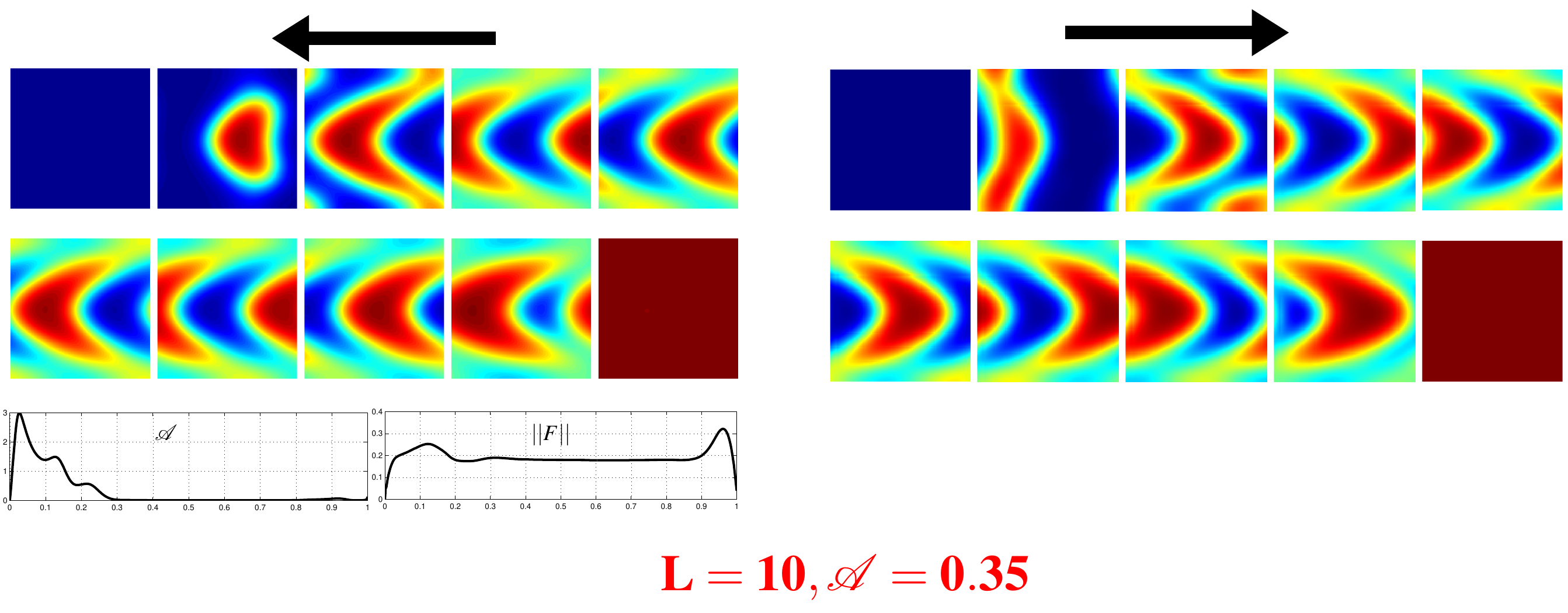}}
\caption{Two different transitions going through two different hyperbolic limit cycles. Same
setup than Fig. \ref{ZON}. They seem to be related to the limit cycle obtained
in \cite{tao} Fig. 13 by a symmetry-breaking mechanism.}
\label{arrows}
\end{figure}

There is another solution with lower action which breaks the translational 
symmetry w.r.t. $x$. It was first detected in \cite{Hey_VdE_prl} and is shown in Fig. \ref{SFI}. To obtain
this solution easily, we start from random initial conditions but
use a ``symmetry breaker" term by adding the penalty constraint in (\ref{generalC}):
\begin{equation}\label{symmbreak}
\gamma_{\rm sym} \left(\Int_0^1 ||\partial_x {\cal U}||^2~d\tau - {\rm cst} \right)^2.
\end{equation}
The NN is forced to find solutions which are not invariant by $x$. The penalty coefficient $\gamma_{\rm sym}$ 
is then set to zero to verify that the solution obtained is stable for the NN representation. 
\\\\
In addition, deep gMAM is also able to detect solutions going through some hyperbolic limit cycles
in the separatrix and obtained by imposing a total length (\ref{instleng}) $L=10$. 
There are shown in Fig. \ref{arrows}. We do not explore further the action functional space, since
it is very rich. It illustrates that deep gMAM is a very powerful tool for detecting various types
of local minimizers. 

\begin{figure}[!h]
\centerline{
\includegraphics[width=9cm]{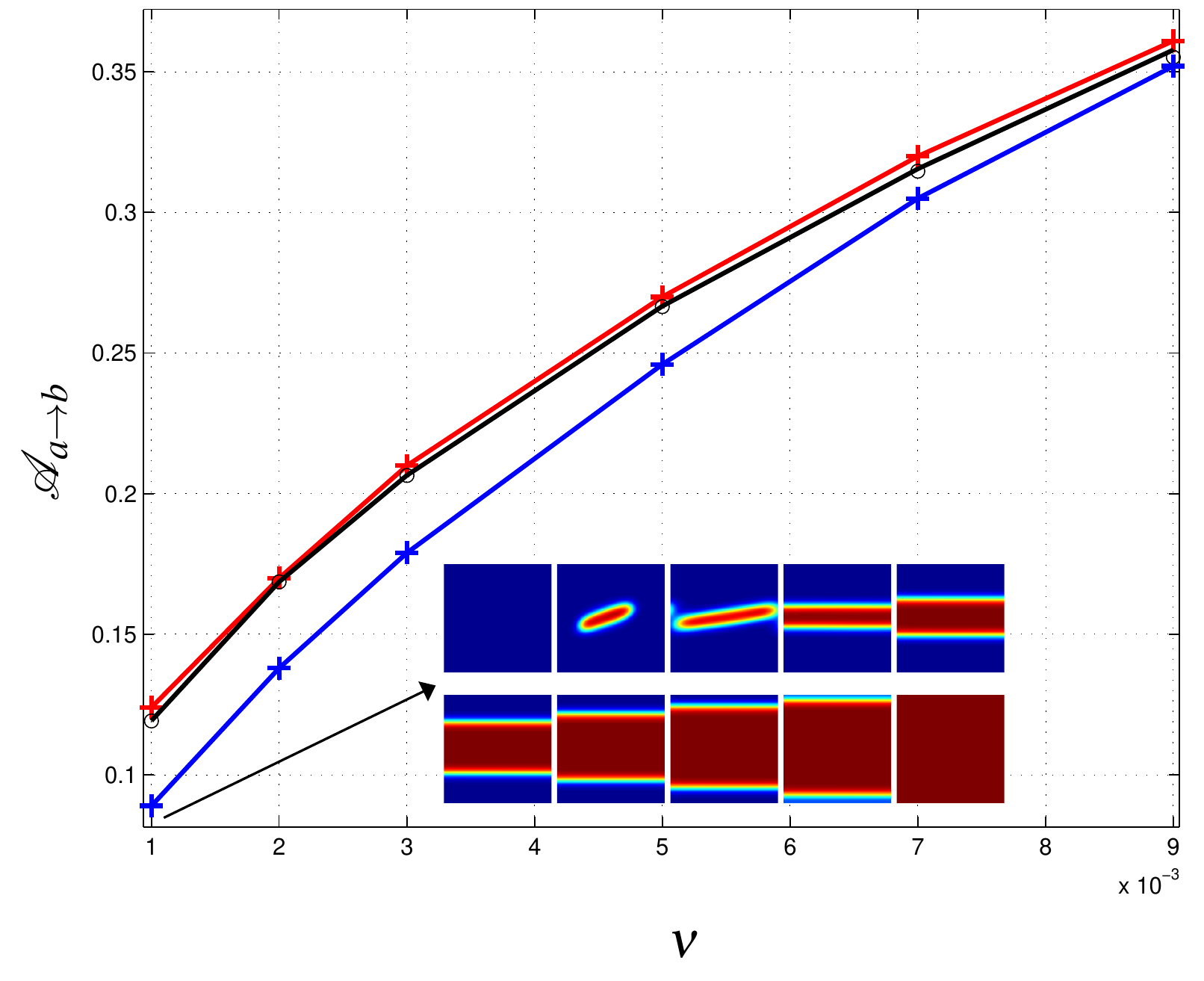}
}
\caption{Dependence of the Arrhenius law with the viscosity $\nu$ for the $x$-invariant solutions of Fig. \ref{ZON} (red curve) compared with the theoretical formula (\ref{arrshear}) (black curve). The instanton solutions which break the $x$ invariance with smaller action
are also shown (blue curve) together with a panel at very low viscosity $\nu = 10^{-3}$ showing the oumuamua typical shape.}
\label{VISC_ARR}
\end{figure}

We conclude this part, by illustrating the use of adiabatic descent \ref{adiad}.
We would like to study the Arrhenius law for Eqs. (\ref{2dshear}) 
as a function of the diffusion coefficient $\nu$. Once the solution shown in Fig. \ref{SFI} is computed,
we can apply the procedure described in (\ref{adiad}) using $\nu$. The boundary layer size decreases
like $\sqrt{\nu}$, it is therefore better to use a logarithmic change, e.g.
$\nu_k = \max(\nu_{\rm end}, \nu_0 (\nu_{\rm end}/\nu_0)^{k/N})$ with $N$ large, we use here $\nu_0 = 9
\cdot 10^{-3}$ and $\nu_{\rm end} = 10^{-3}$.
The result is shown in Fig. \ref{VISC_ARR} using the same NN used in Fig. \ref{ZON}.
It is possible to decrease even further $\nu$ and obtain more precise curves by using NNs with larger capacity.

\subsection{Nongradient case: $d=3$ Sheared Ginzburg-Landau}\label{ng3PDEs}
\begin{figure}[!h]
\centerline{
\includegraphics[width=15cm]{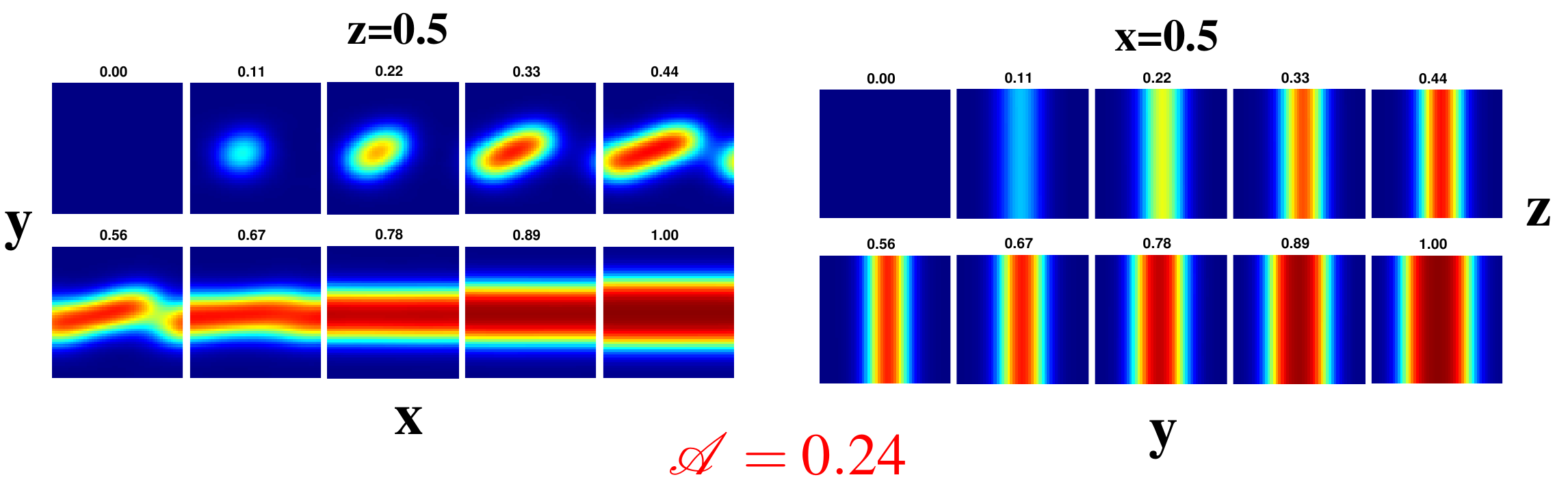}
}
\caption{Codim-1 nucleation obtained using a NN with 7 hidden layers, 10 neurons/layer and
tensorial batches $N_\tau \times N_x = 16 \times 32$ (see (\ref{batch}) and two
symmetry-breakers in $x$ and $y$ (see (\ref{symmbreak})).}
\label{SFI3d}
\end{figure}

The last example illustrates the fact that higher-dimensional PDEs can be considered as well.
In many instances, one wishes to study the action functional space in higher dimensions, say $d=O(10)$
(we recall that it is the dimension of the physical space here).
The case $d=3$ is often encountered in practice, say in hydrodynamics, but should not be considered as an
upper limit of what can be achieved. In fact, one can also increase the input dimension
by adding problem parameters inside the NN parametrization and compute directly a whole family
of instantons. This approach is not used here but an explicit 
example is given in Appendix \ref{Julia}.
A very rough estimate of the typical size 
of the minimisation problem using classical methods when $d=3$ is at best $O(10M)$ dofs, since
the fields are four-dimensional ($d+1$). This is a straightforward consequence of the curse of dimensionality.
In deep gMAM the cost is reduced drastically, the experiment
shown in Fig. \ref{SFI3d} use 1662 parameters only.  The explanation for such an impressive difference,
is in fact that the transition paths have a rather simple structure embedded in a very high-dimensional (phase) space.
In other words, the geometrical structures live in some reduced space of finite and small
dimension. Nevertheless, the ability for simple
NNs to detect complicated embedding is simply outstanding. Figure \ref{SFI3d} shows a typical codim-1
nucleation for the 3-D ACGL PDE (\ref{2dshear}) using the shear $0.1 \sin 2\pi y (\partial_x u + \partial_z u)$
instead. Note that we were not able to find codim-2 nucleations, in this case: all 
codim-2 candidates indeed are relaxing on codim-1 nucleations.

\section{Extreme events in Burgers 1-D}\label{burg1}
The last example is very different from the previous ones. The problems before 
are relying on FW large deviation estimates  of the transition probability between
two bimodal states $a$ and $b$. The small large deviation parameter is the noise amplitude
and one is looking at {\it rare events}.

In this section, we rather
consider some large deviation estimates of the occurrence
of extreme events in Burgers fluid turbulence. The large deviation parameter is
now related to the notion of extreme events instead, here the pointwise
gradient amplitude of the solution denoted $\alpha$ in the following.
Moreover, the problem now involves a more complicated covariance operator $\chi$ so that one 
must use (\ref{aux}). Moreover, 
the ending state $b$ is now replaced by a set of states characterized by an observable.

Burgers equations are very useful for studying turbulence as they share similarities
with the 3-D Navier-Stokes equations, in particular they both have a direct energy cascade
and exhibit {\it intermittency}. This phenomenon is related to the anomalous scaling in
the velocity increments. In more details the mean quantities $S_n = 
\langle |u(x+h) - u(x-h)|^n \rangle$ called structure functions scale like $h^{\xi_n}$ with
some $\xi_n$ smaller than the typical Kolmogorov dimension estimate $n/3$. Closely related
to these quantities are the velocity gradients which is the focus here. A detailed
description of this problem can be found in \cite{Grafke}.
The stochastically-forced 1-D Burgers reads after proper rescaling of $\alpha$  
\begin{equation}
\partial_t u - F(u) = \sigma \eta,~F(u) = -u \partial_x u +  \partial^2_{xx},
\end{equation}
with some additive noise having correlation in space and white in time
$$
\langle \eta(t,x) \eta(t',x') \rangle = \delta(t-t') \chi(x-x'),
$$
and with small amplitude in the limit of extreme events $\alpha_0 \gg 1$
(see Remark 1 in \cite{GYT} and Appendix \ref{Bresca})
$$
\sigma = \left(\frac{\chi_0}{\nu} \right)^\frac12 \frac{1}{\alpha_0}.
$$
One has therefore a coincidence between the limit of small noise amplitude and the limit
of extreme events. Traditionally, one chooses
\begin{equation}
\chi(x) = (1-x^2) {\rm e}^{-\frac{x^2}{2}} = -\partial_{xx}^2 ({\rm e}^{-\frac{x^2}{2}}),
\end{equation}
in order to force the large scales at length $L=1$ (and imposes the solution
to have zero mean). The small scales are thus essentially governed
by the Burgers dynamics through the direct energy cascade.
We are interested in the probability to observe a given velocity gradient at a particular
region of space time, say $t=0,x=0$:
\begin{equation}\label{pra}
\Pr((\partial_x u)(0,0) = \alpha),
\end{equation}
and its behavior when $|\alpha| \to \infty$. The phenomenology is very different depending on the
sign of $\alpha$. As explained in \cite{Balkovsky,Grafke}, for $\alpha > 0$ (right tail), 
the dynamics essentially follow
the characteristics which damp the gradients. The noise must compensate this nonlinear effect
in a significant way so that one expects an important deviation from Gaussianity. 
It can be shown \cite{Gurarie} that in this case $\log \Pr \sim -c_\chi |\alpha|^3$.
The case $\alpha < 0$ (left tail) is more complicated and 
this is the situation we are interested in.
In this case, one has the appearance of shocks regularized by the viscosity $\nu$.
As $|\alpha|$ becomes large, the noise must compensate the viscosity effects preventing the blow up
of the solution. A LDP can be shown to exist in this regime, ie in the viscous tail of the
distribution, with a scaling behavior which can described by the instanton of the FW action.
In such a case, the scaling is known to behave
as \cite{Balkovsky}
\begin{equation}\label{sca32}
\log \Pr \sim -c_{L,\nu,\chi} |\alpha|^\frac32.
\end{equation}
We now describe the minimum action FW problem. Although the following is
technical, it illustrates well how to adapt a complicated problem to a 
setup which works. 
Due to the presence of the covariance term,
the norms involve the convolution operator with $\chi$: 
$$
||f||_\chi = \langle F,\chi \star F \rangle,~ \chi \star F = f.
$$
The geometrical action is now (\ref{finalgeoa}) with the above norm. The initial state $a$
is zero and there is no final state $b$, it is replaced instead by the pointwise 
gradient constraint:
$$
(\partial_x u)(1,0) = \alpha. 
$$
The main technical difficulty is to handle the convolution operator
properly, ie one must solve $\chi \star v = \dot u$ and $\chi \star w = F(u)$ and
use the auxiliary fields $v,w$ in the geometrical action (see Eqs \ref{pengeoa},\ref{aux}). 
It turns out that an efficient
way to do it is to use the Fourier space (continuous) representation, with the unitary conventions
$\hat f(\xi) = (2\pi)^{-\frac12} \Int_{\field{R}} f(x) {\rm e}^{-i \xi x}~dx$, 
$f(x) = (2\pi)^{-\frac12} \Int_{\field{R}} \hat f(\xi) {\rm e}^{i \xi x}~d\xi$. In this
case, we know that (Plancherel)
$$
||f||_\chi = \langle F,\chi \star F \rangle = \langle \hat F, \hat \chi \hat F \rangle = 
||\hat F||_{\hat \chi}.
$$
where the last term, is a simple weighted $L_2$ norm with weight
$$
\hat \chi(\xi) = \xi^2 {\rm e}^{-\frac{\xi^2}{2}}.
$$
The geometrical action is thus
$$
{\cal A}_g = \Int_0^1 (||\hat v||_{\hat \chi} ||\hat w||_{\hat \chi} -
\langle \hat v,\hat w \rangle_{\hat \chi} )~d\tau,
$$
with the two constraints
$$
\hat \chi \hat v - \widehat{\dot u} = 0,~\hat \chi \hat w - \widehat{F(u)} = 0.
$$
The final technical difficulty is to estimate correctly the Fourier transforms $\widehat{\dot u}$
and $\widehat{F(u)}$ using ad-hoc rescaling. 
One can in principle truncate the spatial integrals by relying of
the fast decay of $u$: $u(\tau,x) \to 0$, $x \to \infty$. We do instead a nonlinear transformation
by mapping $\field{R} \to (-1,1)$ using the transformation
$$
Y: x \mapsto \tanh \frac{x}{l}, ~l > 0.
$$
We give the change of variables for convenience, it is
$\Int_{\field{R}} F(u) {\rm e}^{-i \xi x}~dx = \Int_{-1}^{+1} (-u \partial_Y u + 
\nu l^{-1} ((1-Y^2) \partial_{YY}^2 u - 2 Y \partial_Y u)~{\rm exp}(-i \xi l 
{\rm atanh} Y)~dY$. Since we are interested in the case $|\alpha| \to \infty$, we
also need to rescale the problem. Although for moderately large $|\alpha|$ it is not needed, 
it is better to do so to avoid possible NN saturations when the gradients
become too large. We therefore use a NN ansatz following (\ref{ansatz}) except for the
ending condition and with an additional
scaling factor. The ansatz is, using the same notation $u$ in the $Y$ coordinate,
$$
u(\tau,Y) = l^{-1} |\alpha|^\gamma (1-Y^2) \tau {\cal U}(\tau,Y), ~\tau \in [0,1],~
Y \in ]-1,1[.
$$
where ${\cal U}$ is now a feedforward NN. The term $1-Y^2$ not only enables one to 
automatically satisfy 
the boundary conditions at $x \to \pm \infty$ (ie $Y \to \pm 1$) but also avoid 
asymptotical pathological behavior in the NN representation due to the term $\widehat{\dot u}$.
The scaling $l^{-1} |\alpha|^\gamma$ is here to compensate the 
very large amplitude of the instantons.
It can be removed from the equations by rescaling the auxiliary fields $\hat v,\hat w$ as well:
$$
\hat v = l^{-1} |\alpha|^\gamma \hat V,~\hat w  = l^{-1} |\alpha|^\gamma \hat W.
$$
Finally, the penalisation constraints for the auxiliary fields $\hat V,\hat W$ are
after simplifications
$$
{\cal C}_v = \Int_0^1 \Int_0^{\xi_c} \left| \hat \chi \hat V(\xi) - 
 \Int_{-1}^{+1} \partial_\tau (\tau {\cal U}(\tau,Y)) ~{\rm exp}(
(-i \xi l {\rm atanh} Y)~dY 
\right|^2~d\xi d\tau
$$
and
$$
{\cal C}_w = \Int_0^1 \Int_0^{\xi_c} \left| \hat \chi \hat W(\xi) -
 \Int_{-1}^{+1} {\cal F}({\cal U}) ~{\rm exp}(
(-i \xi l {\rm atanh} Y)~dY
\right|^2~d\xi d\tau,
$$
with
$$
{\cal F}({\cal U}) = 
-|\alpha|^\gamma {\cal U}_s \partial_Y {\cal U}_s +
\nu l^{-1} ((1-Y^2) \partial_{YY}^2 {\cal U}_s - 2 Y \partial_Y {\cal U}_s),
~~{\cal U}_s = (1-Y^2) \tau {\cal U}(\tau,Y) .
$$
The integrals in Fourier space are half-integrals using a truncated frequency $\xi_c$
since $u$ is real. The geometrical action is 
\begin{equation}\label{compact}
	{\cal A}_g = l^{-2} |\alpha|^{2\gamma} \Int_0^1 (||\hat V||_{\hat \chi} 
||\hat W||_{\hat \chi} - \langle \hat V,\hat W \rangle_{\hat \chi}) ~d\tau,
\end{equation}
where $\langle \hat V,\hat W \rangle_{\hat \chi} = \int_0^{\xi_c} \Re \{ \hat V 
\overline{\hat W}\} \hat \chi d\xi$.
The choice of $\gamma$ is in principle arbitrary, but according to (\ref{sca32}), it
is convenient to choose $\gamma = \frac34$ so that the integral in
(\ref{compact}) must asymptotically behave independently of $\alpha$.
We thus call this term the {\it compensated geometrical action}.
The gradient constraint in the new $Y$ coordinate after simplifications reads
$$
\partial_Y {\cal U}(1,0)  = l \alpha |\alpha|^{-\gamma}. 
$$
This constraint is handled through a straightforward
quadratic penalisation with a large penalty coefficient. Due to the
rescaling used, the right-hand side term scales like $|\alpha|^{1/4}$ so that large gradient
constraints can be used.
We use a value of $l = 25 $
in applications, a too small 
value would yield a bad representation of the $u$ asymptotical tails $|x| \to \infty$
by strongly zooming into the interval $|x| \ll 1$. 
This setting is likely not optimal (e.g. a better tuning would use $l = l(\alpha)$)
but it works well in practice for $\alpha$ in the range $O(10^2)$--$O(10^3)$.
The auxiliary fields $V,W$ are replaced by two NNs of dimension output 2, one for the
real part and one for the imaginary part. One can also use four NNs instead.
The batches are of two kinds, the ones in physical space $(t,x) \in (0,1) \times (-1,1)$
and the others in the Fourier space $(t,\xi) \in (0,1) \times (0,\xi_c)$ using uniform
distributions.
\\
We show first the instanton path obtained for a small value of $\alpha$ together 
with its spatial profile for $\tau = 1$. Its space-time structure compares 
very well with known results, e.g. Fig. 2 in \cite{Grafke0}. It has zero spatial mean as expected and
shares some similarities with the derivative of the kink solution in Burgers although
with heavier tails and a change of sign visible at $|x|\approx \pm 5$. 
The increase of the distance between the two extrema for $\tau < 1$ is also typical of
the space-time instantons found in the literature.
\begin{figure}[!h]
\centerline{
\includegraphics[width=12cm]{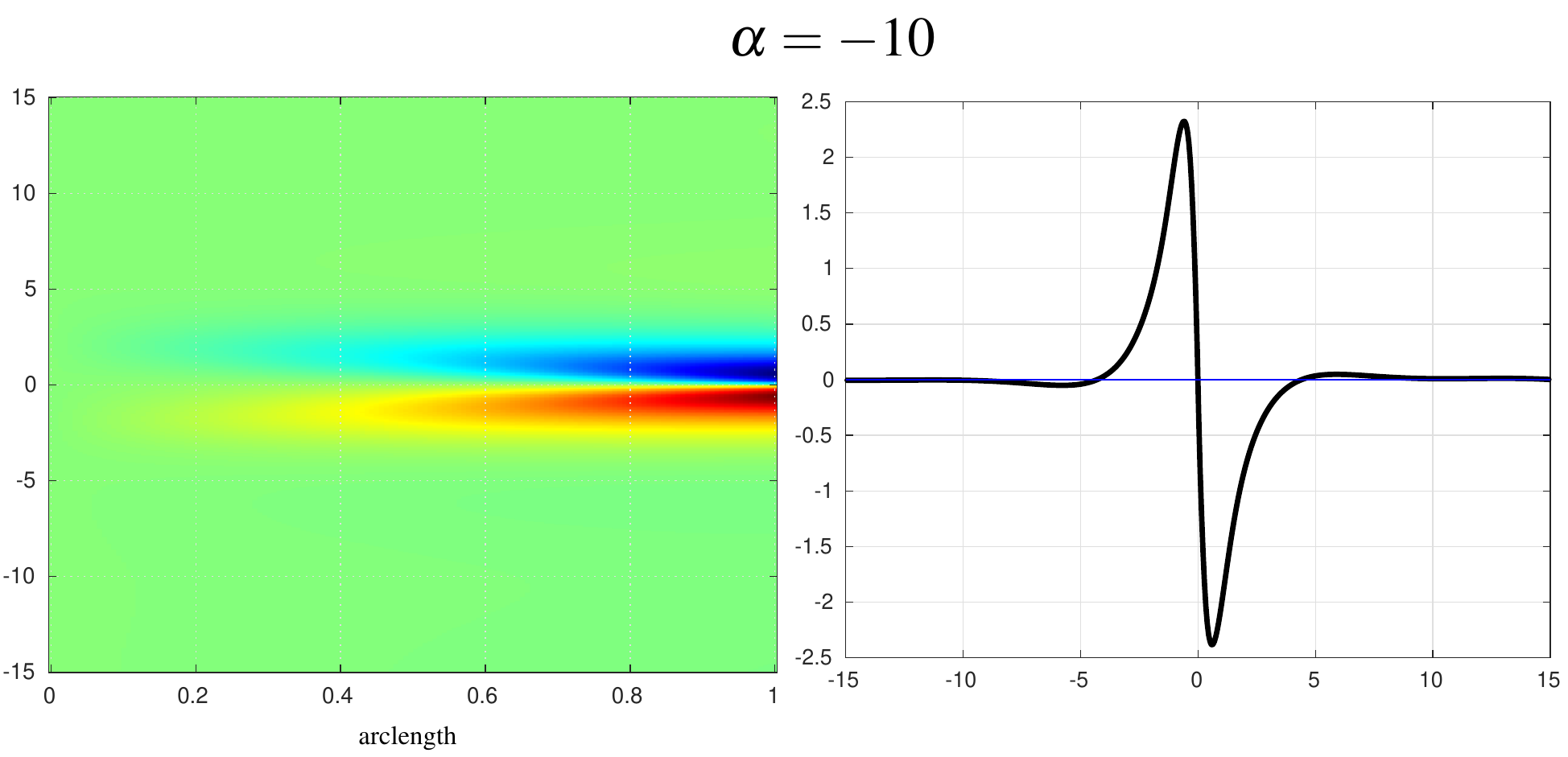}
}
\caption{Instanton solution obtained by deep gMAM using a NN with 14 hidden layers, 10 neurons/layer
and swish activations with batches $N_\tau=15$, $N_Y = 500$, $N_\xi = 500$, training rate
$10^{-4}$, $l = 10$ and $\xi_c = 30$.}
\label{nninst}
\end{figure}
We now show the estimate of the left tail of the distribution by changing the amplitude of the
gradient $\alpha$ to larger values using the setup described above. The method strikingly
recovers the correct exponent $\frac32$. We do not use the adiabatic approach here but rather
a poor-man version using initial conditions from previously computed instantons and freeze
$\alpha$ in the estimates.
The geometrical action estimates are the mean over the last 50~000 iterations. The setup
proposed here is likely not optimal due to the strongly ill-conditioned problem with about 
5\% relative fluctuations from the mean action.
The result shows that the power-law estimate has
in fact a smaller exponent in the range $|\alpha| = O(10^2)$
in accordance with the $1.15$ exponent found in \cite{TVDE,Gotoh} . This can be seen in
the monotonic decrease of the compensated action in the insert of Fig. \ref{powerlaw}.
However, in our case, we are able to explore
an additional decade with very clear convergence to $\frac32$ consistent with the 
theory \cite{Balkovsky}.
\begin{figure}[!h]
\centerline{
\includegraphics[width=11cm]{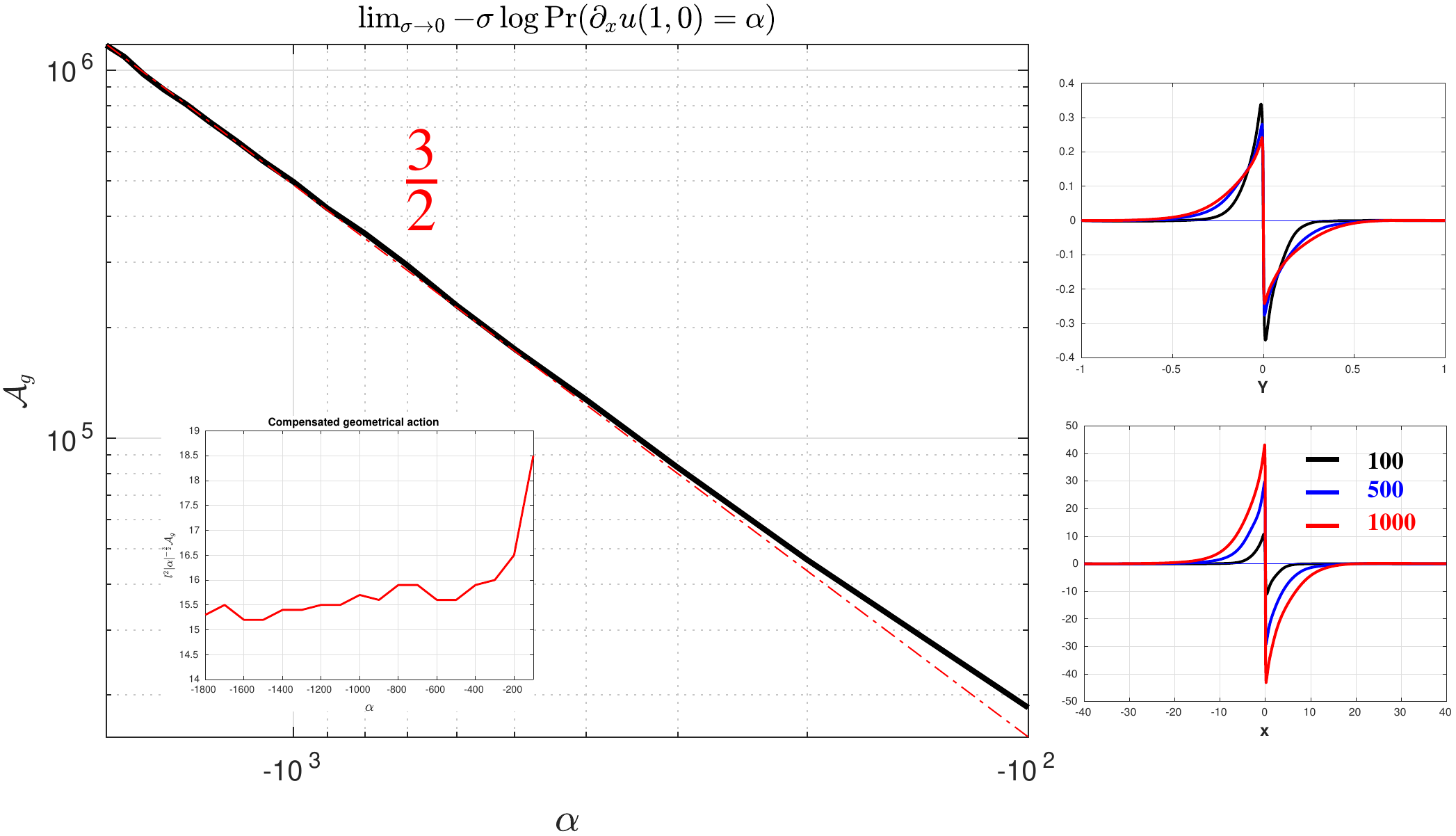}
}
\caption{Power law for the viscous left tail in Burgers shocks. The insert is what is
actually computed, ie the compensated
action as a function of $\alpha$ and the loglog figure is (\ref{compact}) 
(the exact $\frac32$ power law is represented by a dash-dot red line). 
The right panels show the solutions $u(1,Y)$ for different values of
$\alpha$. Upper-right panel are the renormalized curves in $Y$ coordinate and
lower-right panel shows the true ones in $x$ coordinate. The penalisation parameters are
$(\gamma_g,\gamma_{\rm arc},\gamma_{\alpha},\gamma_{v},\gamma_{w}) = (5, 1, 200, 1000, 1000)$
and $l=25$, other parameters like in Fig. \ref{nninst}.}
\label{powerlaw}
\end{figure}
\\
An important remark is that these approaches can be extended to higher-dimensional problems:
the use of the continuous Fourier representation, including the change of variables, together 
with the Monte-Carlo sampling strategy make this possible. This would
be impossible if one attempts to project the fields on a discrete Fourier basis, or
mimicking some FFT representation or any other Galerkin projection, since one would face
very heavy optimisation problems (curse of dimensionality at $d \geq 3$).


\section{Conclusion}\label{Concl}
The deep gMAM method is a natural formulation of the classical gMAM method in the
context of machine learning. It is a penalisation approach
which offers great flexibility at the cost of less precision than classical methods.
However, its outstanding property is its expressivity and its scaling with the dimension, the method
already starts to shine for spatial dimension $d=2$. The reason lies on the well-known 
property for deep networks to overcome the curse of dimensionality.
We have shown various examples, from classical bimodal switches in SDEs and
Ginzburg-Landau PDEs to genuine nongradient, nontransverse situations.
One can easily obtain general Arrhenius laws into a wide variety of contexts.
Furthermore, the deep gMAM approach is able to compute extreme events as well: it is
illustrated in a challenging example of Burgers turbulence. We
are able to recover the correct power law exponent $\frac32$ for the left probability tail at
very small numerical cost.

We discuss below first an important alternative to the FW  minimum action principle and
second some general perspectives.
\subsection{Other approaches: rare event algorithms}\label{AMS}
An important alternative class of methods to estimate rare events are the so-called 
importance sampling \cite{Touchette2} 
and importance splitting methods. The splitting approach is particularly illustrative
of methods which do not exploit some minimum action principle at all. It relies
on smart sampling strategies to provide events
in the tail of the distribution. One can already identify an important advantage compared
to minimum action methods: they can be used in a wider variety
of situations since no LDP is required. In particular, these approaches can consider
finite-amplitude noise rather that the zero noise limit. 
This would be indeed an ideal situation. Unfortunately, there are some fundamental issues which cannot be ignored.  The case
of importance splitting methods is particularly interesting to this respect.  They
originally come from an idea of J.Von Neumann in Los Alamos and were called {\it
Von Neumann splitting} \cite{KH,RR}. A revival of this approach appeared much 
later with the work of mathematicians in probability theory \cite{PdM,Guya_C1} and are now
called adaptive multilevel splitting (AMS) or rare events or even genetic algorithms.
The idea is to draw an ensemble of short simulations and select only the
"most interesting" events as new initial conditions:
this is often referred to as a selection-mutation step owning to some Darwinian selection 
principle in the model.  The main issue is to define what "most interesting" means. These
methods therefore rely on a so-called {\it reaction coordinate} to decide when some event
is better than another. The mutation step is also strongly tightened to the amplitude of the noise
in the model as one must bring diversity 
to obtain relevant statistical ensembles. A too small noise amplitude (or 
a nonchaotic deterministic system) would yield fast species extinction. The method in these
cases are failing to obtain relevant ensembles associated to the tail distribution
unless one uses prohibitively large-sized ensembles (the number of clones).
These methods have thus many issues in the small noise-amplitude limit. The main reason
takes its root in the choice of a good reaction coordinate. The better the reaction coordinate
the less variance in the estimates with nearly Gaussian behavior \cite{Guya_C2,EScomb}
(see also \cite{Brehier} for the general case). 
The less noise, the more sensitive is the choice of the reaction coordinate 
\cite{Joran15,Brehier}. 
It turns out that
the optimal reaction coordinate is known, it is called the committor function and is
a solution of the backward Focker-Planck (Kolmogorov) equation (BFP) \cite{TPT}. 
Since for a system with $N$ d.o.f.s the corresponding BFP
equations is a $N$-dimensional PDEs, it is therefore a very challenging if not an impossible task
unless $N=O(10$-$100)$ at most (for models in fluid dynamics, $N=O(10^{6-7})$ and more).
In fact, being able to compute a pointwise estimate
of the committor is just what these rare events algorithms are trying to do.
There are lots of works devoted to this question. In the machine-learning context,
one can indeed try to compute the committor function in some reduced space of
collective variables \cite{Lelievre2}, (see also \cite{khoo2019solving}). 
A more involved approach denoted loop-AMS has been proposed recently which amounts to identify
a better reaction coordinate in a dynamical way \cite{FBanalog}.
Although the results are encouraging, they
also face the issue of the small noise-amplitude limit: the committor function in 
phase space becomes nearly singular 
with Heaviside-type of behavior. The corresponding 
exponentially flat regions are indeed reflecting the large deviation scaling
of the probability to estimate.
One can also consider spectral theory \cite{Professors,Schutte} to investigate the BFP spectral
gap and leading spectrum but again
the singular small-amplitude limit  \cite{gaspard2005chaos} cannot be handled easily.
Nevertheless, AMS approaches using non-optimal reaction coordinates with large enough
noise-level show spectacular results which cannot be ignored: recent examples
include heat waves 
\cite{RAGONE:2018:A}, 1-D Ginzburg-Landau \cite{Joran}
and quasigeostrophic 2-D turbulence \cite{PRLme,JASme}, transition to turbulence 
\cite{Joran22}.

For these reasons, if the noise level in the model is too small
and an effective action is known (here the FW action), it is much more reliable
to consider minimum action methods in general. Both approaches are complementary however:
they are optimal in two almost separated regimes.

\subsection{Perspectives}
The next immediate perspective is to consider situations where there is no access
to an explicit Lagrangian, see Eqs (\ref{Ham0}). In this case, one must
solve the minimax problem using adversarial networks by some 
stochastic descent-ascent algorithms. Although the general ideas of the deep gMAM
method are the same, it requires a more specific and careful treatment with ad-hoc examples.

In general, one would like to extend the domain of validity of the proposed approach.
First, an important theoretical question is to extend the method beyond the FW regime.
Having access to an explicit macroscopic action is not always possible, even worst
the existence of a LDP is not guaranteed at all.
An illuminating example is the 2-D quasigeostrophic
barotropic equation stochastically forced at the small scales. It is used as a model
for mid-latitude atmospheric and oceanic jets through an inverse turbulent cascade (zonostrophy,
see \cite{JASme}). In such a case,
the FW action related to bimodal switches is not relevant anymore \cite{JLaurie}. One needs
an effective action which can be identified by advanced kinetic analysis \cite{Freddy1,Tomas}.
In this scenario, the use of rare events algorithms such
as AMS (see subsection \ref{AMS}) seems unavoidable and gives good results as soon as the noise
amplitude is large enough \cite{PRLme}.

Another situation is the important case of chaotic deterministic systems. Very often,
there is no LDP and the mean waiting time is in fact replaced by some power law \cite{Grebo2}.
We discuss in the Appendix 
\ref{FWzero} a striking example of a simple chaotic system for reversal dynamo which has
both. The use of deep gMAM reveals that the FW small action minimizers
are strongly connected if not equal to the exact preferred transition path.
We thus believe that deep gMAM can be extended in much more general situations, much beyond the LDP hypothesis. It also suggests that a marriage between splitting methods and gMAM approaches is
indeed possible.

\section*{Acknowledgments}
The author thanks T.Grafke, E.Vanden-Eijnden, R.Zakine and M.D.Chekroun for stimulating discussions and support.

\section{Appendix}

\subsection{Bimodality: minimizing the FW zero action paths}\label{FWzero}
In many cases, the addition of noise 
in a deterministic system -- often for modelisation reasons -- is necessary to
observe transitions between two attractors $a$ and $b$. One calls 
them noise-induced since the purely deterministic system is not able by itself to 
trigger transitions between the attracting states.
In the context of chaotic deterministic systems, the situation can be very different.
Chaos can often triggers transitions between $a$ and $b$ through 
intermittency alone: the transitions are no more noise-induced.
As soon as, a timescale separation is present, these transitions are readily
interpreted as rare events. However, their probabilities of occurrence do not satisfy some LDP.
The example is a simple model of dynamo reversal having three d.o.fs \cite{Giss,G2}:
\begin{equation}\label{G}
\dot x = \mu x - yz,\dot y = -\nu y + x z, \dot z = -z  + xy + \Gamma.
\end{equation}

\begin{figure}[!h]
\centerline{
\includegraphics[width=7cm]{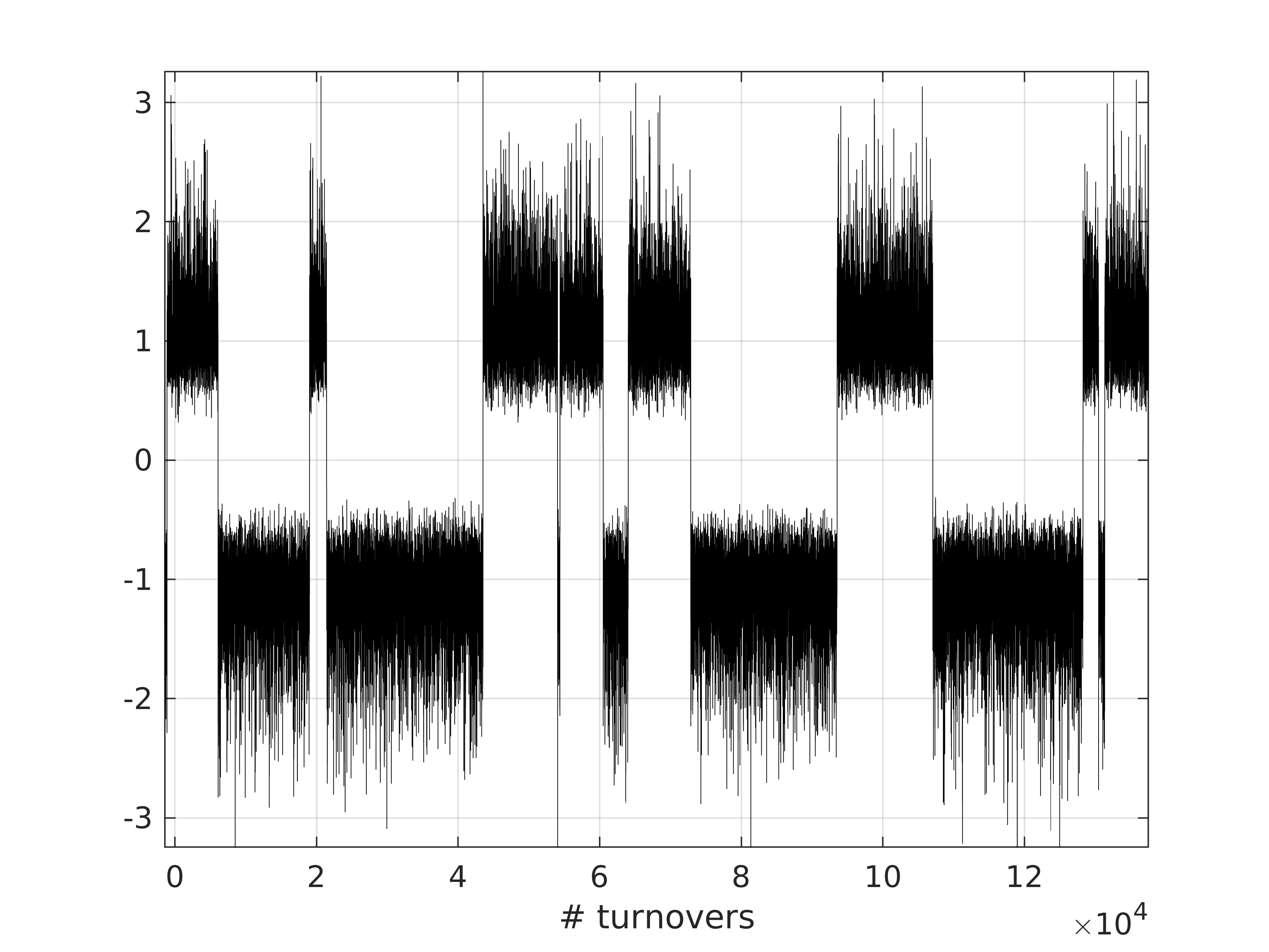}
\includegraphics[width=7cm]{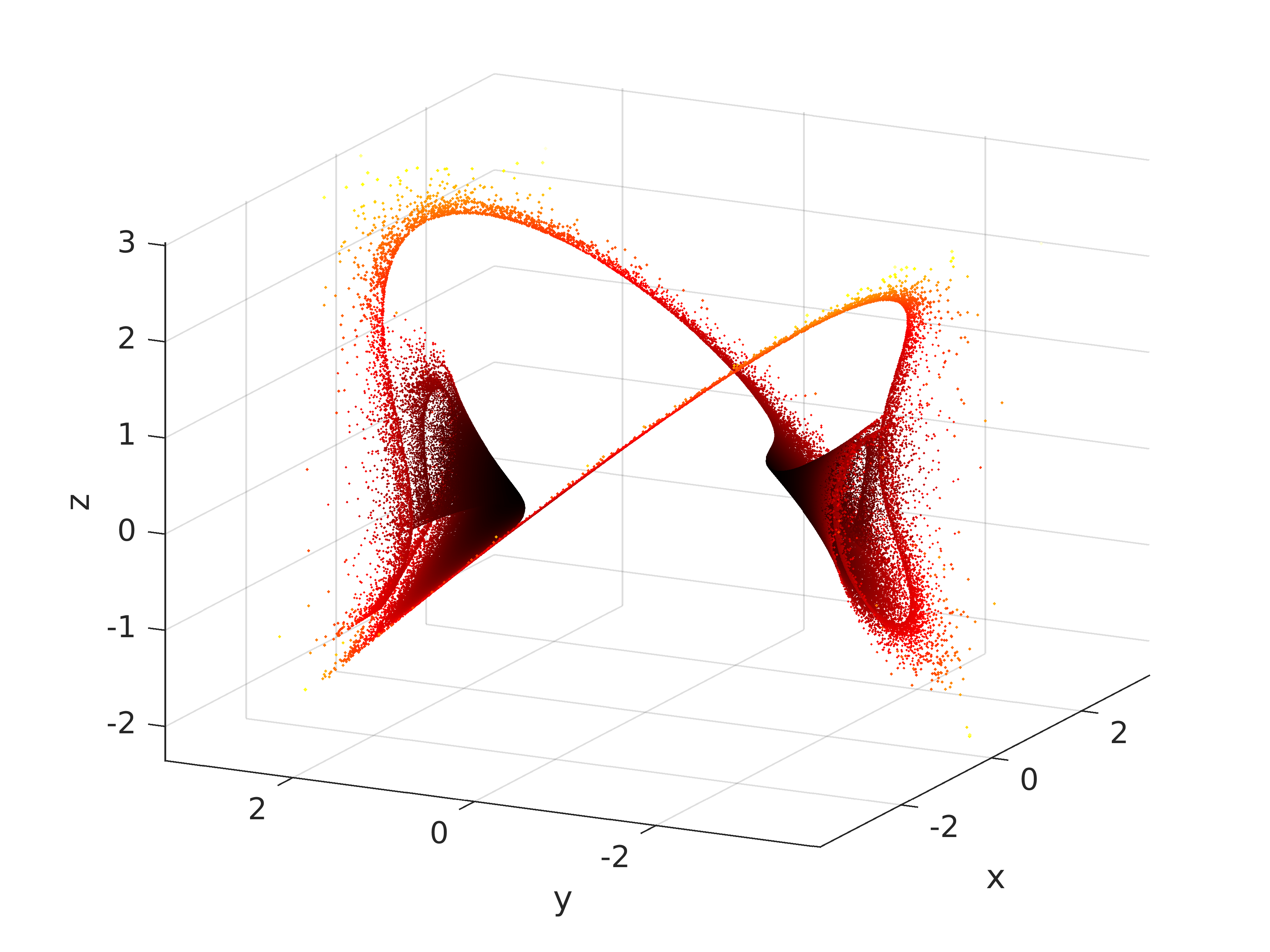}
}
\caption{Bimodality in system (\ref{G}): timeseries (left panel) versus attractor in phase space (right panel) using $\mu = 0.1192805,\nu=0.1,\Gamma = 0.9$.}
\end{figure}
Such phenomenon is well-known in chaos theory and is called {\it crisis-induced intermittency},
see the seminal work of Grebogi, Ott and York \cite{Grebo}. 
It occurs when two disconnected basins of attraction 
start to overlap at a given critical value of the system parameters. 
In chaotic systems having discrete group symmetries such scenario is the rule rather than the exception
\cite{Chossat}. It appears that the mean waiting time $\tau$, ie the mean time interval between two
transitions is scaling like a power law rather than exponentially:
$$
\tau \sim (\mu_c-\mu)^{-\gamma},~\mu < \mu_c
$$
with $\gamma \sim \frac12$ in the dynamo model (\ref{G}) \cite{G2}. In fact, the critical exponent
depends on the type of crisis tangency \cite{Grebo2}.
With this in mind, one can ask about the status of the FW action (see (\ref{FWA}) with
$\chi = Id$). Since no noise is
actually needed to observe a transition, the transition paths are purely deterministic.
The action along these paths is therefore identically zero.

\begin{figure}[!h]
\centerline{
\includegraphics[width=15cm]{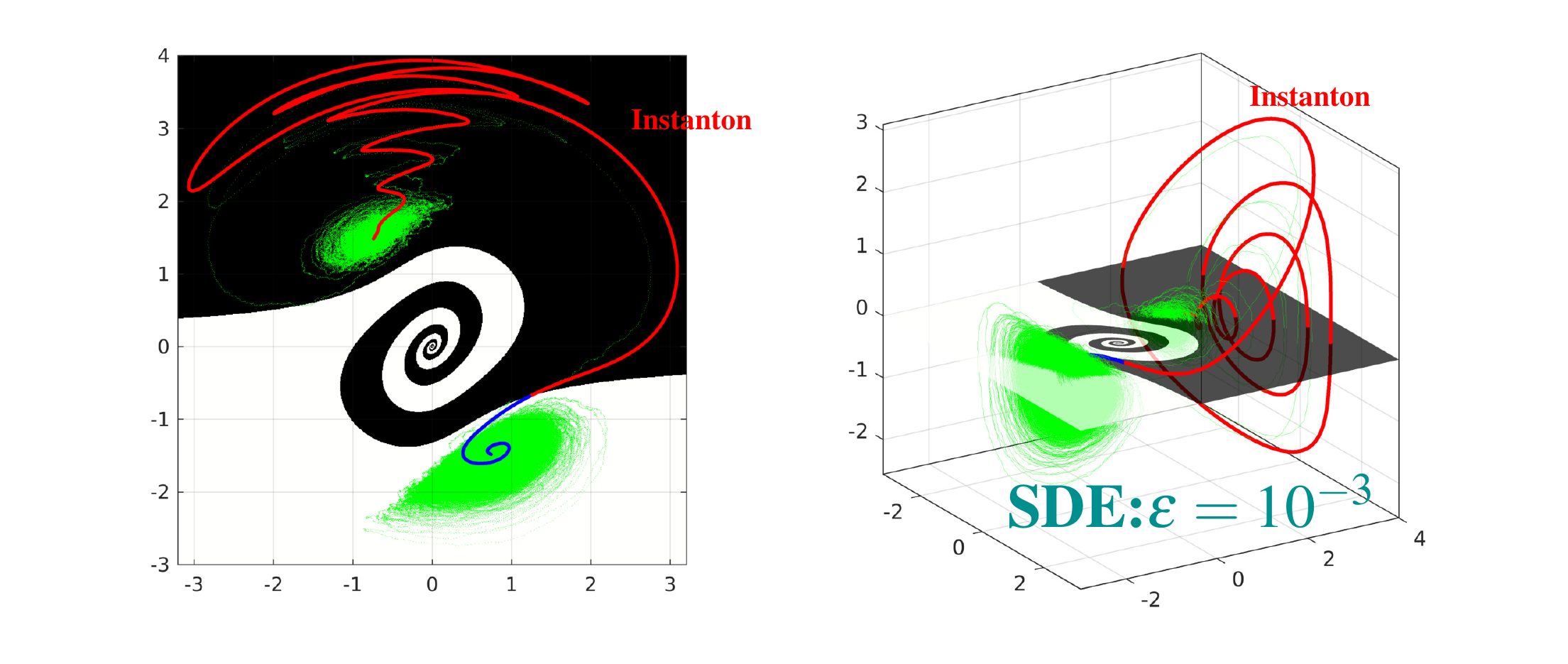}
}
\caption{Instanton path computed by deep gMAM in the FW regime $(\mu > \mu_c)$.
The black and white spiraling regions show the oscillatory separatrix around $O$ and the green
curves are stochastic direct simulations with small noise $\epsilon$.}
\label{BYU}
\end{figure}

Before the crisis bifurcation ($\mu > \mu_c$), it is clear however that the 
transitions must be noise-induced and FW theorem does apply. Let us look in more details.
The system has one critical point $O = (0,0,\Gamma)$ and two stable oscillatory points
at $S_\pm \equiv (\pm (\nu + \Gamma (\nu/\mu)^\frac12)^\frac12,-\pm 
(\nu + \Gamma (\mu/\nu)^\frac12)^\frac12,-(\mu \nu)^\frac12)$.
An important particularity is that $O$ has Morse index 2 with an oscillatory unstable manifold.
In this case, the instanton path does not go through $O$ but crosses the separatrix through
a nontrivial point. This contrasts with the Lorenz63 system \cite{Lorenz63} where transition 
paths do cross the separatrix through the saddle \cite{Zhou}.
It shows that for nongradient systems, 
critical points in the separatrix are not guaranteed to play 
a role. The nature of these points in the separatrix is the subject
of many theoretical researches, they are often called {\it noisy precursor of crisis}
\cite{Silchenko,Kraut,Demaeyer}. The instanton path is shown in Fig. \ref{BYU} 
using deep gMAM. It agrees with  direct ensemble simulations of (\ref{G}) with a small additive Gaussian noise (not shown).

\begin{figure}[!h]
\centerline{
\includegraphics[width=14cm]{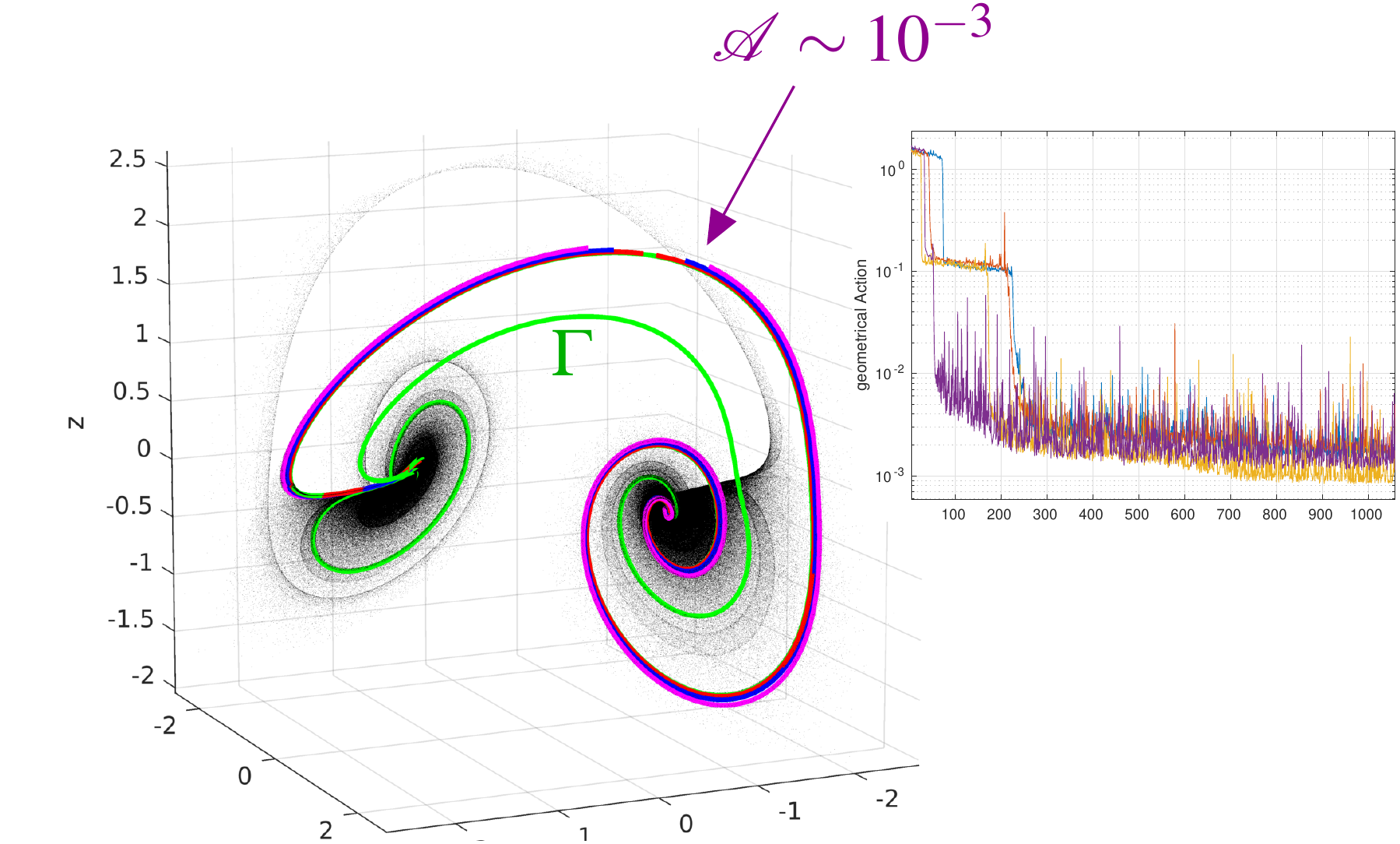}
}
\caption{Near zero-action instantons paths with different initial conditions $a$ using deep gMAM:
trajectories in phase space (left panel) and action during the
stochastic gradient descent . The red-purple curves concentrate on the deterministic transition
whereas the outlier green curve is correct up to a scaling factor.}
\label{goodtr}
\end{figure}

We now consider the regime where FW do not apply, ie in the crisis regime for $\mu < \mu_c$.
We naively apply deep gMAM on various initial conditions on the attractor $a$.
The result is striking and is shown in Fig. \ref{goodtr}.
First the instanton paths not surprisingly depend on the initial condition,
second and more importantly, many of them do agree with the deterministic transition
by concentrating around it (red-blue curves in Fig. \ref{goodtr}). 
The green curve is an outlier with a transition path which 
differs from the exact solution. However, 
this outlier is indeed correct up to a scaling factor: there is some $g > 0$ such that $g\Gamma$
is in the neighborhood of the deterministic transition.

In principle it would be interesting to relax the BV constraints using instead 
a whole set of conditions in the neighborhood of the attracting points $S_\pm$.
For instance, by penalizing the characteristic functions 
$\chi_{B(S_\pm,\rho)}$ where $\chi_{\cal D} = 1$ if ${\bf x} \in {\cal D}$ and zero otherwise,
like in section  \ref{burg1}. In fact, AMS also uses a 
similar idea by sampling the equilibrium measure on a codim-1 surface surrounding the attractors
\cite{Lelievre1}. It appears however that the correct approach can be more subtil,
it is discussed in \cite{Kraut}.

One of interpretation for the validity of FW transition paths in the near zero-action regime
is mostly related to the use of the geometrical action itself and should not come as a surprise. 
Instead of integrating the
system for a very long time, one is able to detect trajectories with time reparametrization
by minimisation of the geometrical action. It therefore bypasses 
the long residence time (and accumulation points) for the state to visit the attractors.
We believe that such phenomenon should be analyzed in more details  on more complicated systems
as it is potentially a very strong alternative to rare events algorithms even when no LDP
exists. In such scenario, one has no more access to probability estimates since the LDP interpretation has disappeared but nevertheless the correct transitions are captured. 
Probability estimates must therefore be obtained by other approaches
like AMS (see subsection \ref{AMS}).

\subsection{A simple Julia code with comments}\label{Julia}
We illustrate here the great flexibility, simplicity
and efficiency of neural networks in general when coupled with 
user-friendly modern languages (Julia,
PyTorch). We provide here a snippet Julia code for computing the quasi-potential
in subsection \ref{msqv}: it is short, simple and can be easily modified and extended.
We do not write the monitoring and display subroutines as they are user-dependent.
It is important for instance to test different hyperparameters (NN size, training rate, batch
size, activation functions, descent algorithm, etc...).
One can also set \texttt{din = 1} for computing only one instanton. In this case, it is interesting to add the arclength condition (\ref{arc}). One must simply add \verb!mean(A2) - (mean(A))^2!
to the returned quantity in the function loss, with some (small) penalty weight.
Note that when $N_\tau > 1$, the arclength penalty is no more additive.
One can also explore different
ODEs, e.g. Lorenz63, or even setting \texttt{din = 5} for computing $(a,b) \mapsto V(a,b)$.
A more difficult exercise is then to switch to the PDE case.

\begin{figure}[!h]
\centerline{
\includegraphics[width=11cm]{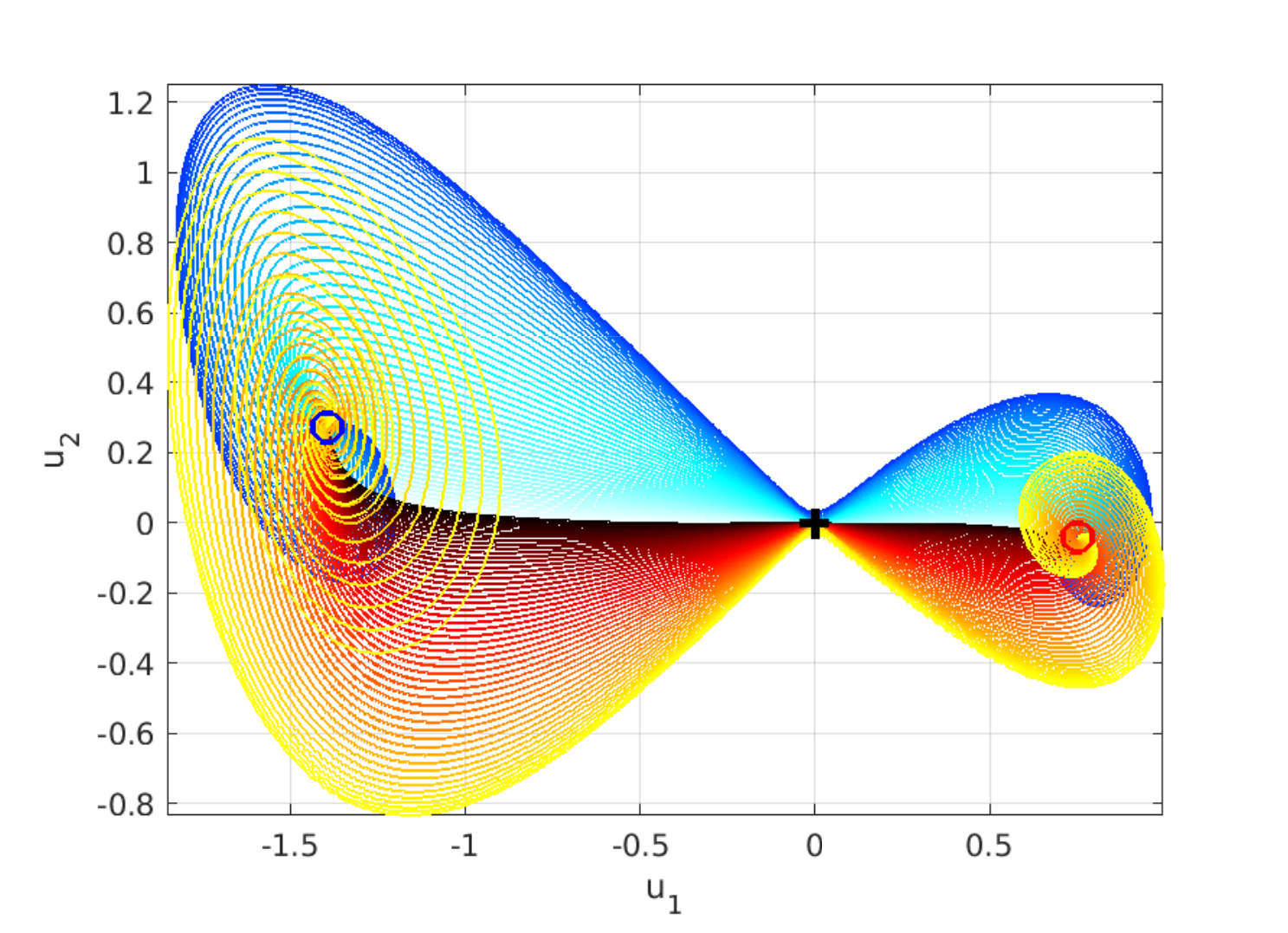}
}
\caption{Family of instantons for $a \to b$ (red colors) and 
$b \to a$ (blue colors). They all cross the saddle (black cross) and are spiralling
outwards or towards $a$(left) and $b$(right) in the Hamiltonian regime when $\sigma \to 0$. 
The dark-colored instantons correspond to the gradient regime ($\sigma \sim 1$).
Note that in this example the transitions $b \to a$ are much more likely than $a \to b$
and can be quantified precisely by monitoring the action values. }
\label{fig8}
\end{figure}

We give here another simple example of a 2 d.o.fs family of SDEs depending
on some parameter $\sigma$ such that it is gradient when $\sigma=1$ and
Hamiltonian when $\sigma = 0$. Let $u = (u_1,u_2)$ and
\begin{equation}\label{family}
F(u,\sigma) = \sigma G + (1-\sigma) G^\perp,
G = (-u_1^3 + u_1 - 0.3 u_1^2 u_2 - 0.6 u_1^2,-u_2 - 0.1 u_1^3).
\end{equation}
By definition this system has the transverse decomposition (\ref{transverse}) and 
classical Arrhenius laws can be obtained explicitly \cite{tao}.
It has two stable points and a saddle point at the origin. When $\sigma < 1$, 
the geometrical support of the
transition $a \to b$ differs from $b \to a$: it is called the figure-eight scenario. 
The transitions $a \to b$ must all cross the saddle $(0,0)$ and then relax on deterministic zero-action paths spiraling towards $b$ (and outwards from $a$). 
The cost functional is $ {\cal C}[u] = \Int_0^1 d\sigma \Int_0^1 d\tau A_g(u(\tau,\sigma))$.
Figure \ref{fig8}  gives the typical result obtained
for the two transitions $a \to b$ and $b \to a$. The code is essentially the same
code than below with very few modifications.
\\
\begin{verbnobox}[\mbox{}\footnotesize]
#~~~~ Compute the Maier-Stein quasi-potential b->V(a,b) for a=(-1,0)
using Flux,Distributions
#~~~~~~~~~~~~~~~~~~~~~~~~~~~~~~~~~~~~~~~~~~~~~~~~~~
N    = 2                              # Maier-Stein has only N=2 dofs
din  = N+1                            # first dim is tau dim 2,...,N are for b
ee   = vcat(1e-6,zeros(N))            # fd tau increment [eps,0,...,0]
a    = [-1;0]                         # ODE fixed pts a (N=2 only)
cap  = 15                             # NN capacity neurons/layer
act  = swish                          # activation function
tr   = 1e-3                           # training rate
NT   = 1000                           # batch size
Nmax = 50000                          # max numb of descent iterations
opt  = Flux.ADAM(tr)                  # we use ADAM descent with training rate tr
#~~~~~~~~~~~~~~~~~~~~~~~~~~~~~~~~~~~~~~~~~~~~~~~~~~
# We define a feedforward NN with input dimension N+1 and output dimension N for (u1,...,uN)
# it has 2 hidden layers having cap neurons/layer using swish activations 
# This NN is likely too small -> one must increase the # of hidden layers 
		
NN   = Chain(Dense(din,cap,act),Dense(cap,cap,act),Dense(cap,cap,act),Dense(cap,N))
ps   = Flux.params(NN)                # define the NN parameters

#~~~~~~~~~~~~~~~~~~~~~~~~~~~~~~~~~~~~~~~~~~~~~~~~~~
#  the BV ansatz: U(0) = a, U(1) = b, see Eq.(13)
function U(r)
 tau  = r[1:1,:]
 b    = r[2:din,:]
 mtau = 1.0 .- tau
 return a*mtau + b.*tau + tau.*mtau.*NN(r)
end
#~~~~~~~~~~~~~~~~~~~~~~~~~~~~~~~~~~~~~~~~~~~~~~~~~~
#  Maier-Stein system with N = 2 d.o.fs 
function RHS(u)
 u1 = u[1:1,:]
 u2 = u[2:2,:]
 F1 = u1 - u1.^3 -10*u1.*u2.^2
 F2 = -(1.0 .+ u1.^2).*u2
 return [F1;F2]
end
#~~~~~~~~~~~~~~~~~~~~~~~~~~~~~~~~~~~~~~~~~~~~~~~~~~
function loss(r)                         # r is a random set of pts of size (din,NT)
 dotU    = (U(r.+ee)-U(r.-ee))/(2*ee[1]) # or using Zygote AD
 FU      = RHS(U(r))                 
 A2      = sum(abs2,dotU,dims=1)         # L2 norm squared ||dotU||^2(tau,b)
 A       = sqrt.(A2)                     # L2 norm ||dotU||(tau,b)
 B       = sqrt.(sum(abs2,FU,dims=1))    # L2 norm ||F(U)||(tau,b) 
 AB      = sum(dotU.*FU,dims=1)          # Hilbert product <dotU,F(U)>(tau,b) 
 ACTION  = mean(A.*B) - mean(AB)         # sum over b and tau: see Eq.(25)
 return  ACTION 
end
#~~~~~~~~~~~~~~~~~~~~~~~~~~~~~~~~~~~~~~~~~~~~~~~~~~
# The stochastic gradient descent using ADAM to find all the instantons
for k in 1:Nmax
 batch = rand(din,NT)                  # generate a total of NT random pts for (tau,b) in (0,1)^3
 gs = gradient(ps) do                  # Zygote AD of the cost functional
   loss(batch)                         # Zygote AD of the cost functional
 end                                   # Zygote AD of the cost functional
 Flux.update!(opt,ps,gs)               # NN parameters update using ADAM
end
#~~~~~~~~~~~~~~~~~~~~~~~~~~~~~~~~~~~~~~~~~~~~~~~~~~
\end{verbnobox}

\subsection{Burgers rescaling}\label{Bresca}
We set $u = U u'$, $x = L x'$, $t = T t'$. 
The noise term therefore scales as $\eta = (\chi_0^\frac12 T^{-\frac12}) \eta'$. 
The extreme event condition reads $(U/L) (\partial_{x'} u')(1,0) = \alpha_0 \alpha'$ and
we take $\alpha_0 = U/L$. Choosing $U = L/T$, $T=L^2/\nu$ gives $L = (\nu/\alpha_0)^\frac12$
and
$$
\partial_t u + u \partial_x u - \partial^2_{xx} u = \sigma \eta
~{\rm with}~
\sigma = \left(\frac{\chi_0}{\nu}\right)^\frac12 \frac{1}{\alpha_0} 
\xrightarrow[\alpha_0 \to \infty]{} 0
$$
One can therefore use the FW formalism in this context \cite{TVDE,GYT}. When $\nu$ is small, one expects
that the asymptotical regime is attained for much larger $\alpha_0$. This is self-consistent
with the idea that atypical extreme events must be much more extreme when the Reynolds is large
than the ones in the small Reynolds number regime.
\bibliographystyle{plain}
\bibliography{references}

\begin{thebibliography}{10}

\bibitem{Balkovsky}
E.~Balkovsky, G.~Falkovich, I.~Kolokolov, and V.~Lebedev.
\newblock Intermittency of burgers' turbulence.
\newblock {\em Phys. Rev. Lett.}, 78:1452--5, 1997.

\bibitem{beck2021deep}
Ch. Beck, S.~Becker, P.~Cheridito, A.~Jentzen, and A.~Neufeld.
\newblock {Deep splitting method for parabolic PDEs}.
\newblock {\em SIAM J. Sci. Comp.}, 43(5):A3135--A3154, 2021.

\bibitem{Lelievre2}
Z.~Belkacemi, G.~Paraskevi, T.~Leli\`evre, and G.~Stoltz.
\newblock Chasing collective variables using autoencoders and biased
  trajectories.
\newblock {\em arXiv:2104.11061v2}, 2022.

\bibitem{Freddy1}
F.~Bouchet, C.~Nardini, and T.~Tangarife.
\newblock {Kinetic theory of jet dynamics in the stochastic barotropic and 2D
  Navier-Stokes equations}.
\newblock {\em J. Stat. Phys.}, 153(4):572--625, 2013.

\bibitem{PRLme}
F.~Bouchet, J.~Rolland, and E.~Simonnet.
\newblock Rare event algorithm links transitions in turbulent flows with
  activated nucleations.
\newblock {\em Phys. Rev. Lett.}, 122(7):074502, 2019.

\bibitem{Brehier}
C.E. Br\'ehier, M.~Gazeau, L.~Goudenege, T.~Leli\`evre, and M.~Rousset.
\newblock Unbiasedness of some generalized adaptive multilevel splitting
  algorithms.
\newblock {\em Annals of Appl. Prob.}, 26(6):3559--3601, 2016.

\bibitem{Cai1}
W.~Cai and Zhi-Qin~J. Xu.
\newblock Multi-scale deep neural networks for solving high dimensional {PDEs}.
\newblock {\em arXiv:1910.11710v1}, 2019.

\bibitem{Cameron}
M.K. Cameron.
\newblock {Finding the quasipotential for nongradient SDEs}.
\newblock {\em Physica D: Nonlinear Phenomena}, 241(18):1532--1550, 2012.

\bibitem{OJK}
A.~Cavagna, A.J. Bray, and Rui~D.M. Travasso.
\newblock {Ohta-Jasnow-Kawasaki} approximation for nonconserved coarsening
  under shear.
\newblock {\em Phys. Rev. E}, 62:4702--4719, Oct 2000.

\bibitem{Guya_C1}
F.~C\'erou and A.~Guyader.
\newblock {Adaptive Multilevel Splitting for Rare Events Analysis}.
\newblock {\em Stoch. Anal. and Appl.}, 25:417--443, 2007.

\bibitem{Lelievre1}
F.~C\'erou, A.~Guyader, T.~Leli\`evre, and D.~Pommier.
\newblock A multiple replica approach to simulate reactive trajectories.
\newblock {\em J.Chem.Phys.}, 134(054108), 2011.

\bibitem{qmc}
J.~Chen, R.~Du, P.~Li, and L~Lyu.
\newblock Quasi-monte carlo sampling for solving partial differential equations
  by deep neural networks.
\newblock {\em Numer. Math. Theor. Meth. Appl.}, 2020.

\bibitem{Chossat}
P.~Chossat and M.~Golubitsky.
\newblock Symmetry-increasing bifurcation of chaotic attractors.
\newblock {\em Physica D}, 88:423--436, 1988.

\bibitem{Demaeyer}
J.~Demaeyer and P.~Gaspard.
\newblock A trace formula for activated escape in noisy maps.
\newblock {\em J. Stat. Mech.: Theor. and Exp.}, 2013(10):P10026, 2013.

\bibitem{periodic_dong}
S.~Dong and N.~Ni.
\newblock A method for representing periodic functions and enforcing exactly
  periodic boundary conditions with deep neural networks.
\newblock {\em J. Comp. Phys.}, 435:110242, 2021.

\bibitem{EVdE}
W.~E, W.~Ren, and E.~Vanden-Eijden.
\newblock {Minimum action method for the study of rare events}.
\newblock {\em Comm. Pure Appl. Math.}, 57:1--20, 2004.

\bibitem{TPT}
W.~E and E.~Vanden-Eijnden.
\newblock Towards a theory of transition paths.
\newblock {\em J. Stat. Phys.}, 123:503--523, 2006.

\bibitem{Jona}
W.G. Faris and G.~Jona-Lasinio.
\newblock Large fluctuations for a nonlinear heat equation with noise.
\newblock {\em J.Phys. A: Math. Gen.}, 15, 1982.

\bibitem{FW}
M.~I. Freidlin and A.~D. Wentzell.
\newblock {\em {Random Perturbations of Dynamical Systems, volume 260 of
  Grundlehren der Mathematischen Wissenschaften}}.
\newblock Springer-Verlag, New York, 1984.

\bibitem{FW98}
M.~I. Freidlin and A.~D. Wentzell.
\newblock {\em {Random Perturbations of Dynamical Systems, 2nd ed., volume 260
  of Grundlehren der Mathematischen Wissenschaften}}.
\newblock Springer-Verlag, New York, 1998.

\bibitem{gaspard2005chaos}
P.~Gaspard.
\newblock {\em Chaos, Scattering and Statistical mechanics}.
\newblock Number~9. Cambridge University Press, 2005.

\bibitem{G2}
C.~Gissinger.
\newblock A new deterministic model for chaotic reversals.
\newblock {\em Eur. Phys. J. B}, 85(137), 2012.

\bibitem{Giss}
C.~Gissinger, E.~Dormy, and S.~Fauve.
\newblock Morphology of field reversals in turbulent dynamos.
\newblock {\em Euro. Phys. Lett.}, 90:49001, 2010.

\bibitem{xavier}
X.~Glorot and Y.~Bengio.
\newblock Understanding the difficulty of training deep feedforward neural
  networks.
\newblock {\em Aistats}, 9:249--256, 2010.

\bibitem{Gotoh}
T.~Gotoh.
\newblock {Probability density dunctions in steady-state Burgers turbulence}.
\newblock {\em Phys. Fluids}, 11:2143--2148, 1999.

\bibitem{Grafke0}
T.~Grafke, R.~Grauer, and T.~Sch\"afer.
\newblock {Instanton filtering for the stochastic Burgers equation}.
\newblock {\em J. Phys. A: Math. Theor.}, 46:062002, 2013.

\bibitem{Grafke}
T.~Grafke, R.~Grauer, and T.~Sch\"afer.
\newblock The instanton method and its numerical implementation in fluid
  mechanics.
\newblock {\em J. Phys. A: Math. Theor.}, 48:333001, 2015.

\bibitem{TVDE}
T.~Grafke, R.~Grauer, T.~Sch\"afer, and E.~Vanden-Eijnden.
\newblock {Relevance of instantons in Burgers turbulence}.
\newblock {\em EPL}, 109(3):34003, 2015.

\bibitem{Grebo2}
C.~Grebogi, E.~Ott, F.~Romeira, and J.A. York.
\newblock {\em Phys. Rev. A}, 36:5365--5380, 1987.

\bibitem{Grebo}
C.~Grebogi, E.~Ott, and J.A. Yorke.
\newblock Crises, sudden changes in chaotic attractors, and transient chaos.
\newblock {\em Physica D}, 7:181--200, 1983.

\bibitem{Gurarie}
V.~Gurarie and A.~Migdal.
\newblock Instantons in the burgers equation.
\newblock {\em Phys. Rev. E}, 54:4908--14, 1996.

\bibitem{Guya_C2}
A.~Guyader, N.~Hengartner, and E.~Matzner-L$\o$ber.
\newblock Simulation of extreme quantiles and extreme probabilities.
\newblock {\em Appl. Math. Optim.}, 64:171--196, 2011.

\bibitem{Hey_VdE}
M.~Heymann and E.~Vanden-Eijnden.
\newblock The geometric minimum action method: A least action principle on the
  space of curves.
\newblock {\em Comm. Pure Appl. Maths}, 61(8):1052--1117, 2008.

\bibitem{Hey_VdE_prl}
M.~Heymann and E.~Vanden-Eijnden.
\newblock Pathways of maximum likelihood for rare events in nonequilibrium
  systems: Application to nucleation in the presence of shear.
\newblock {\em Phys. Rev. Lett.}, 100:140601, Apr 2008.

\bibitem{Aug}
J.~Huang, H.~Wang, and T.~Zhou.
\newblock An augmented lagrangian deep learning method for variational problems
  with essential boundary conditions.
\newblock {\em Commun. Comput. Phys.}, 31:966--986, 2022.

\bibitem{Touchette2}
Y.~Jiawei, H.~Touchette, and G.M. Rotskoff.
\newblock Learning nonequilibrium control forces to characterize dynamical
  phase transitions.
\newblock {\em Phys. Rev. E}, 105:024115, Feb 2022.

\bibitem{KH}
H.~Kahn and T.~Harris.
\newblock Estimation of particle transmission by random sampling.
\newblock {\em Natl. Bur. Stand. Appl. Math. Ser.}, 12:27--30, 1951.

\bibitem{khoo2019solving}
Y.~Khoo, J.~Lu, and L.~Ying.
\newblock Solving for high-dimensional committor functions using artificial
  neural networks.
\newblock {\em Research in the Mathematical Sciences}, 6(1):1--13, 2019.

\bibitem{adam}
D.~P. Kingma and J.~Ba.
\newblock {ADAM: A method for stochastic optimization}.
\newblock {\em arXiv preprint arXiv:1412.6980}, 2014.

\bibitem{Kraut}
S.~Kraut and C.~Grebogi.
\newblock Escaping from nonhyperbolic chaotic attractors.
\newblock {\em Phys. Rev. Lett.}, 92(23):234101, 2004.

\bibitem{JLaurie}
J.~Laurie and F.~Bouchet.
\newblock Computation of rare transitions in the barotropic quasi-geostrophic
  equations.
\newblock {\em New J. of Phys.}, 17:015009, 2015.

\bibitem{Cai2}
Z.~Liu, W.~Cai, and Zhi-Qin~J. Xu.
\newblock Multi-scale deep neural network (mscalednn) for solving
  poisson-boltzmann equation in complex domains.
\newblock {\em arXiv:2007.11207v3}, 2020.

\bibitem{Lorenz63}
E.N. Lorenz.
\newblock Deterministic nonperiodic flow.
\newblock {\em J. Atmos. Sci.}, 20(2):130--141, 1963.

\bibitem{FBanalog}
D.~Lucente, J.~Rolland, C.~Herbert, and F.~Bouchet.
\newblock Coupling rare event algorithms with data-based learned committor
  functions using the analogue markov chain.
\newblock {\em arXiv:2110.05050v3}, 2022.

\bibitem{Maier}
R.S. Maier and D.L. Stein.
\newblock A scaling theory of bifurcations in the symmetric weak-noise escape
  problem.
\newblock {\em J. Stat. Phys.}, 83(3--4):291357, 1996.

\bibitem{DomDec}
B.~Moseley, A.~Markham, and T.~Nissen-Meyer.
\newblock {Finite basis physics- informed neural networks (FBPINNs): a scalable
  domain decomposition approach for solving differential equations}.
\newblock {\em arXiv:2107.07871}, 2021.

\bibitem{Muller}
K.~M\"uller.
\newblock Reaction paths on multidimensional energy hypersurfaces.
\newblock {\em Angew. Chem.}, 92, 1980.

\bibitem{Onsager}
L.~Onsager and S.~Machlup.
\newblock Fluctuations and irreversible processes: Ii. systems with kinetic
  energy.
\newblock {\em Phys. Rev.}, 91:1512--5, 1953.

\bibitem{PdM}
{P. Del Moral}.
\newblock {\em {Feynman-Kac formulae, Genealogical and interacting particle
  systems with applications}}.
\newblock Probability and Applications, Springer-Verlag, New York, 2004.

\bibitem{poppe}
G.~Poppe and T.~Sch\"afer.
\newblock {Computation of minimum action paths of the stochastic nonlinear
  Schr\"odinger equation with dissipation}.
\newblock {\em J. Phys. A: Math. Theor.}, 51 335102:1--17, 2018.

\bibitem{RAGONE:2018:A}
F.~Ragone, J.~Wouters, and F.~Bouchet.
\newblock {Computation of extreme heat waves in climate models using a large
  deviation algorithm}.
\newblock {\em PNAS}, {115}({1}):{24--29}, {} {2018}.

\bibitem{pinns}
M.~Raissi, P.~Perdikaris, and G.E. Karniadakis.
\newblock Physics-informed neural networks: A deep learning framework for
  solving forward and inverse problems involving nonlinear partial differential
  equations.
\newblock {\em J. Comp. Phys.}, 378:686--707, 2019.

\bibitem{swish}
Prajit Ramachandran, Barret Zoph, and Quoc~V. Le.
\newblock Swish: a self-gated activation function.
\newblock {\em arXiv: Neural and Evolutionary Computing}, 2017.

\bibitem{Joran22}
J.~Rolland.
\newblock Collapse of transitional wall turbulence captured using a rare events
  algorithm.
\newblock {\em J. Fluid Mech.}, 931 A22:503--523, 2022.

\bibitem{Joran}
J.~Rolland, F.~Bouchet, and E.~Simonnet.
\newblock {Computing transition rates for the 1-D stochastic
  Ginzburg--Landau--Allen--Cahn equation for finite-amplitude noise with a rare
  event algorithm }.
\newblock {\em J. Stat. Phys.}, 162:277--311, 2016.

\bibitem{Joran15}
J.~Rolland and E.~Simonnet.
\newblock Statistical behavior of adaptive multilevel splitting algorithms in
  simple models.
\newblock {\em J. Comp. Phys.}, 283:541--558, 2015.

\bibitem{RR}
M.~Rosenbluth and A.~Rosenbluth.
\newblock {Monte Carlo calculation of the average extension of molecular
  chains}.
\newblock {\em J. Chem. Phys.}, 23(2):356--359, 1955.

\bibitem{GYT}
T.~Schorlepp, T.~Grafke, and R.~Grauer.
\newblock {Gel’fand–Yaglom type equations for calculating fluctuations
  around instantons in stochastic systems}.
\newblock {\em J. Phys. A: Math. Theor.}, 54:235003, 2021.

\bibitem{Tobiasns3d}
T.~Schorlepp, T.~Grafke, S.~May, and R.~Grauer.
\newblock Spontaneous symmetry breaking for extreme vorticity and strain in the
  3d navier-stokes equations.
\newblock {\em Phil Trans Roy Soc A}, 380(2226), 2022.

\bibitem{Silchenko}
A.N. Silchenko, S.~Beri, D.G. Luchinsky, and P.V. McClintock.
\newblock Fluctuational transitions through a fractal basin boundary.
\newblock {\em Phys. Rev. Lett.}, 91(17):174104, 2003.

\bibitem{EScomb}
E.~Simonnet.
\newblock Combinatorial analysis of the adaptive last particle method.
\newblock {\em Stats and Comp.}, 26:211--230, 2016.

\bibitem{Professors}
E.~Simonnet and M.D. Chekroun.
\newblock Deep spectral computations in linear and nonlinear diffusion
  problems.
\newblock {\em arXiv:2207.03166v1}, 2022.

\bibitem{JASme}
E.~Simonnet, J.~Rolland, and F.~Bouchet.
\newblock Multistability and rare spontaneous transitions in barotropic
  beta-plane turbulence.
\newblock {\em J. Atmos. Sci.}, 78(6):1889--1911, 2021.

\bibitem{dgm}
J.~Sirignano and K.~Spiliopoulos.
\newblock {DGM: A deep learning algorithm for solving partial differential
  equations}.
\newblock {\em J. Comp. Phys.}, 375:1339--1364, 2018.

\bibitem{Tomas}
T.~Tangarife.
\newblock {\em {Kinetic theory and large deviations for the dynamics of
  geophysical flows}}.
\newblock PhD thesis, ENS de Lyon, ENS Lyon,
  https://tel.archives-ouvertes.fr/tel-01241523, 11 2015.

\bibitem{tao}
M.~Tao.
\newblock Hyperbolic periodic orbits in nongradient systems and
  small-noise-induced metastable transitions.
\newblock {\em Physica D: Nonlinear Phenomena}, 363:1--17, 2018.

\bibitem{Touchette}
H.~Touchette.
\newblock The large deviation approach to statistical mechanics.
\newblock {\em Physics Reports}, 478(1):1--69, 2009.

\bibitem{Mullervde}
E.~Vanden-Eijnden and M.~Heymann.
\newblock The geometric minimum action method for computing minimum energy
  paths.
\newblock {\em J. Chem. Phys.}, 128:061103, 2008.

\bibitem{VdE_geo}
E.~Vanden-Eijnden and M.~Heymann.
\newblock {The geometric minimum action method: A least action principle on the
  space of curves}.
\newblock {\em Comm. on Pure and Applied Math.}, 61(8):1052--1117, 2008.

\bibitem{Schutte}
W.~Zhang, T.~Li, and C.~Sch\"utte.
\newblock Solving eigenvalue {PDEs} of metastable diffusion processes using
  artificial neural networks.
\newblock {\em J. Comput. Phys.}, 465:111377, 2022.

\bibitem{Zhou}
X.~Zhou and W.~E.
\newblock {Study of noise-induced transitions in the Lorenz system using the
  minimum actio method}.
\newblock {\em Commun. Math. Sci.}, 8:341--355, 2010.

\bibitem{NeuralPDE}
K.~Zubov, Z.~McCarthy, Y.~Ma, F.~Calisto, V.~Pagliarino, S.~Azeglio,
  L.~Bottero, E.~Luj\'an, V.~Sulzer, A.~Bharambe, N.~Vinchhi, K.~Balakrishnan,
  D.~Upadhyay, and C.~Rackauckas.
\newblock {NeuralPDE: Automating Physics-Informed Neural Networks (PINNs) with
  Error Approximations}.
\newblock {\em arXiv:2107.09443v1}, 2021.

\end{thebibliography}
\end{document}